\documentclass[final,5p,times,twocolumn]{elsarticle}

\usepackage{graphicx}
\graphicspath{{/figures/}}
\DeclareGraphicsExtensions{{eps}{epsi}}

\usepackage{P0DNIM}

\journal{Nuclear Instruments and Methods in Physics}

\begin{document}

\begin{frontmatter}

\author[56]{S.\,Assylbekov}
\author[66]{G.\,Barr}
\author[56]{B.E.\,Berger}
\author[64]{H.\,Berns}
\author[59]{D.\,Beznosko}
\author[63]{A.\,Bodek}
\author[63]{R.\,Bradford\fnref{fn1}}
\author[56]{N.\,Buchanan}
\author[63]{H.\,Budd}
\author[56]{Y.\,Caffari}
\author[64]{K.\,Connolly}
\author[62]{I.\,Danko}
\author[56]{R.\,Das}
\author[64]{S.\,Davis}
\author[63]{M.\,Day}
\author[62]{S.\,Dytman}
\author[64]{M.\,Dziomba}
\author[63]{R.\,Flight}
\author[64]{D.\,Forbush}
\author[59]{K.\,Gilje}
\author[62]{D.\,Hansen}
\author[59]{J.\,Hignight}
\author[59]{J.\,Imber}
\author[61]{R.A.\,Johnson}
\author[59]{C.K.\,Jung}
\author[56]{V.\,Kravtsov}
\author[59]{P.T.\,Le}
\author[59]{G.D.\,Lopez}
\author[59]{C.J.\,Malafis}
\author[63]{S.\,Manly}
\author[61]{A.D.\,Marino\corref{cor1}}
\ead{alysia.marino@colorado.edu}
\cortext[cor1]{Corresponding author}
\author[63]{K.S.\,McFarland}
\author[59]{C.\,McGrew}
\author[47]{C.\,Metelko}
\author[59]{G.\,Nagashima}
\author[62]{D.\,Naples}
\author[47]{T.C.\,Nicholls}
\author[59]{B.\,Nielsen}
\author[62]{V.\,Paolone}
\author[59]{P.\,Paul}
\author[47]{G.F.\,Pearce}
\author[47]{W.\,Qian}
\author[59]{K.\,Ramos}
\author[56]{E.\,Reinherz-Aronis}
\author[63]{P.A.\,Rodrigues}
\author[56]{D.\,Ruterbories}
\author[59]{J.\,Schmidt}
\author[56]{J.\,Schwehr}
\author[47]{M.\,Siyad}
\author[59]{J.\,Steffens}
\author[59]{A.S.\,Tadepalli}
\author[59]{I.J.\,Taylor}
\author[47]{M.\,Thorpe}
\author[56,59]{W.\,Toki}
\author[61]{C.\,Vanek}
\author[56]{D.\,Warner}
\author[66,47]{A.\,Weber}
\author[64]{R.J.\,Wilkes}
\author[56]{R.J.\,Wilson}
\author[59]{C.\,Yanagisawa\fnref{fn2}}
\author[61]{T.\,Yuan}

\fntext[fn1]{Presently at Argonne National Laboratory, Argonne, Illinois, U.S.A}
\fntext[fn2]{Also at BMCC/CUNY, New York, New York, U.S.A.}

\address[61]{University of Colorado at Boulder, Department of Physics, Boulder, Colorado, U.S.A.}
\address[56]{Colorado State University, Department of Physics, Fort Collins, Colorado, U.S.A.}
\address[59]{State University of New York at Stony Brook, Department of Physics and Astronomy, Stony Brook, New York, U.S.A.}
\address[66]{Oxford University, Department of Physics, Oxford, United Kingdom}
\address[62]{University of Pittsburgh, Department of Physics and Astronomy, Pittsburgh, Pennsylvania, U.S.A.}
\address[63]{University of Rochester, Department of Physics and Astronomy, Rochester, New York, U.S.A.}
\address[47]{STFC, Rutherford Appleton Laboratory, Harwell Oxford, United Kingdom}
\address[64]{University of Washington, Department of Physics, Seattle, Washington, U.S.A.}



\title{The T2K ND280 Off-Axis Pi-Zero Detector}



\begin{abstract}
The pi-zero detector (\pod{}) is one of the subdetectors that makes up the off-axis near detector for the Tokai-to-Kamioka (T2K) long baseline neutrino experiment.    The primary goal for the \pod{} is to measure the relevant cross-sections for neutrino interactions that generate \pizero{}'s, especially the cross-section for neutral current \pizero{} interactions, which are one of the dominant sources of background to the \numu{}\goesto{}\nue{}  appearance signal in T2K.   The \pod{} is composed of layers of plastic scintillator alternating with water bags and brass sheets or lead sheets and is one of the first detectors to use Multi-Pixel Photon Counters (MPPCs) on a large scale.

\end{abstract}


\end{frontmatter}


\section{Introduction}
\label{sec:Introduction} 



\begin{figure}[htbp]
\centering
\includegraphics[width=3.5in]{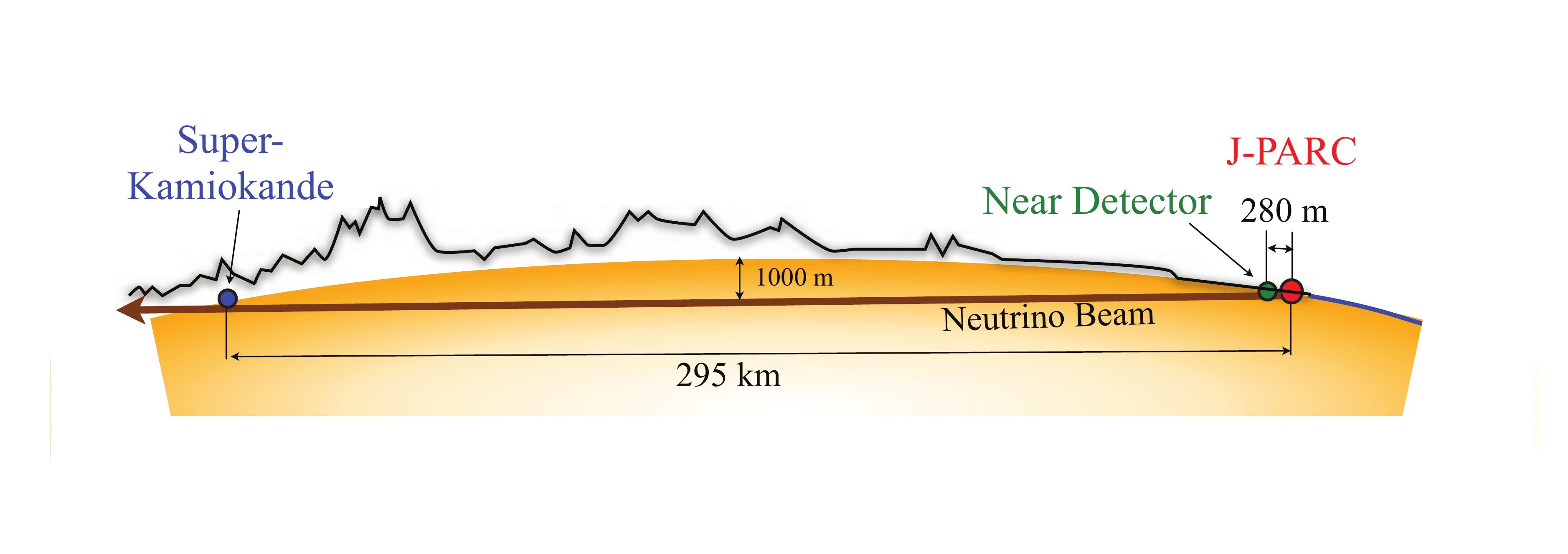}
\caption{A schematic showing the path of the neutrinos in the T2K experiment, from the start of the neutrino beamline at J-PARC to Super-Kamiokande, 295 km away.}
\label{fig:t2k-layout}
\end{figure}

The Tokai-to-Kamioka (T2K) experiment is a long-baseline neutrino oscillation experiment designed to probe the mixing of the muon neutrino with other neutrino species and to shed light on the neutrino mass scale.  The T2K neutrino beam is generated using the the new high-intensity proton  synchrotron at J-PARC, which has a Phase-I design beam power of  0.75~MW.  T2K uses Super-Kamioka\-nde~\cite{SKNIM} as the far detector to measure neutrino rates at a distance of 295~km from the beam production point, and near detectors to sample the unoscillated beam.   The neutrino beam is directed 2.5\degree{} away from the Super-Kamiokande detector  and travels through the Earth's crust under Japan, as illustrated in Fig.~\ref{fig:t2k-layout}.  

The T2K experiment~\cite{T2KNIM} near detector complex (ND280), located 280~m from the start of the neutrino beam, contains the on-axis  INGRID detector and an off-axis detector.

\begin{figure}[htbp]
\centering
\includegraphics[width=3in]{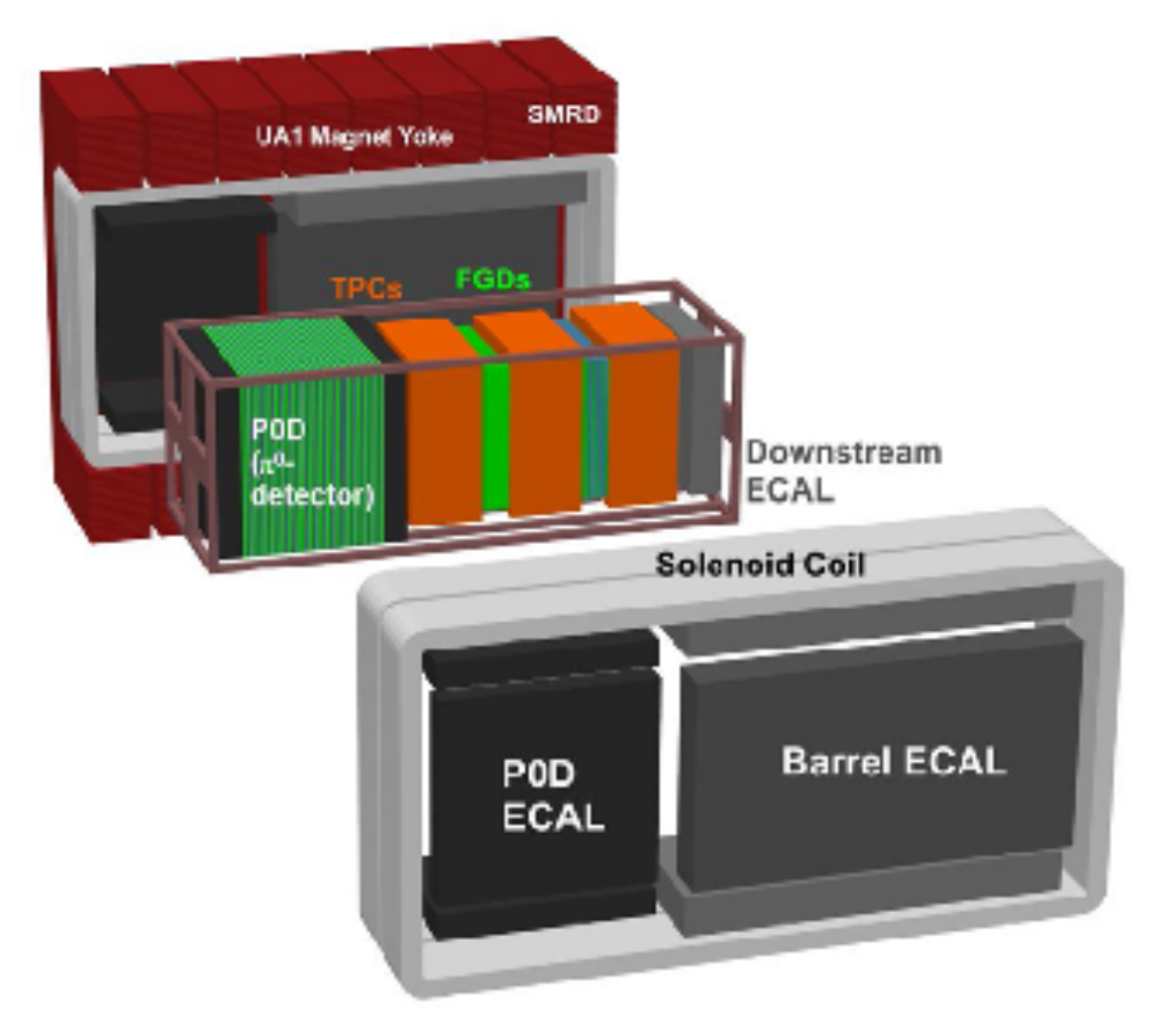}
\caption{An exploded view of the off-axis near detector.}
\label{fig:nd280view}
\end{figure}

The off-axis detector, shown in Fig.~\ref{fig:nd280view}, is situated at the same off-axis angle as Super-Kamiokande and contains the Pi-Zero detector (\pod{}) a plastic scintillator-based detector optimized for
$\pi^0$ detection followed by a tracking detector comprising two fine grained scintillator detector modules (FGDs) sandwiched between three time projection chambers (TPCs).
The \pod{} and tracker are surrounded by electromagnetic calorimeters (ECALs), including a module that sits immediately downstream of the tracker.
The whole detector is located in a magnet with a 0.2~T magnetic field, which also serves as mass for a side muon range detector (SMRD). This paper describes the \pod{}  subdetector in greater detail.

\subsection{Goals of the \pod{}}

The primary physics goal of T2K is to measure the mixing angle $\theta_{13}$ or to improve the existing limit by an order of magnitude if the angle is too small to measure directly. This is done by looking for the appearance of $\nue$ in a $\numu$ beam.  Additional physics goals include the precision determination of the $\dms{}_{23}$ and $\theta_{23}$ parameters through a $\nu_{\mu}$ disappearance measurement, where the parameters will be measured to a precision of $\delta{}(\dms{}_{23}) \sim 10^{-4}\eVs{}$ and $\delta{}(\sstt{}_{23}) \sim 0.01$ respectively.
In addition to neutrino oscillation studies, the T2K neutrino beam (a narrow-band beam with a peak energy of about 600 MeV) will enable a rich physics program of
neutrino interaction studies at energies covering the transition between
the quasi-elastic  and resonance production regimes.

To achieve the required precision for the $\nu_e$ appearance measurement (observed via the process $\nu _{e}+n\rightarrow e^{-}+p$),  the neutral current \pizero{} rate ($\nu _{\mu }+N\rightarrow \nu _{\mu }+N+\pi ^{0}+X$) must be
measured at the J-PARC site near the neutrino beam production point using the off-axis near detector. Events containing $\pi^0$'s are the dominant physics  background to the \nue{} appearance signal at Super-Kamiokande. The \pod{}
sits at the upstream end of the off-axis detector and has been designed
to
precisely measure the neutral
current process $\nu _{\mu }+N\rightarrow \nu _{\mu }+N+\pi ^{0}+X$ on a
water ({\it $H_{2}O$}) target. In addition the \pod{} will constrain the intrinsic $\nu_e$ content of the beam flux which is an irreducible background to the $\nu_e$ appearance measurement.


Early design studies demonstrated that understanding the \pizero{} and
\nue{} backgrounds required sensitivity to interactions containing
\pizero{} with momentum greater than 200\unit{\MeV{}\per{}c}.  This
requires a photon reconstruction threshold of well below 100\unit{\MeV{}}.
Both of the background processes to be constrained by the \pod{} are a
relatively small fraction of the total \pod{} interaction rate, and must be
measured on a water target, forcing a large water mass.  In addition,
sufficient energy resolution is needed to demonstrate the presence of a
\pizero{} through reconstruction the invariant mass.  The eventual design
of the \pod{} realizes these goals by interleaving water target between
scintillator layers which both measure charged particles and support the
water target.  The rate on water is determined using statistical
subtraction with data collected during periods having water in the detector
and out of the detector.

\subsection{Description of the \pod{}}
\begin{figure}[htbp]
\centering
\includegraphics[scale=.3]{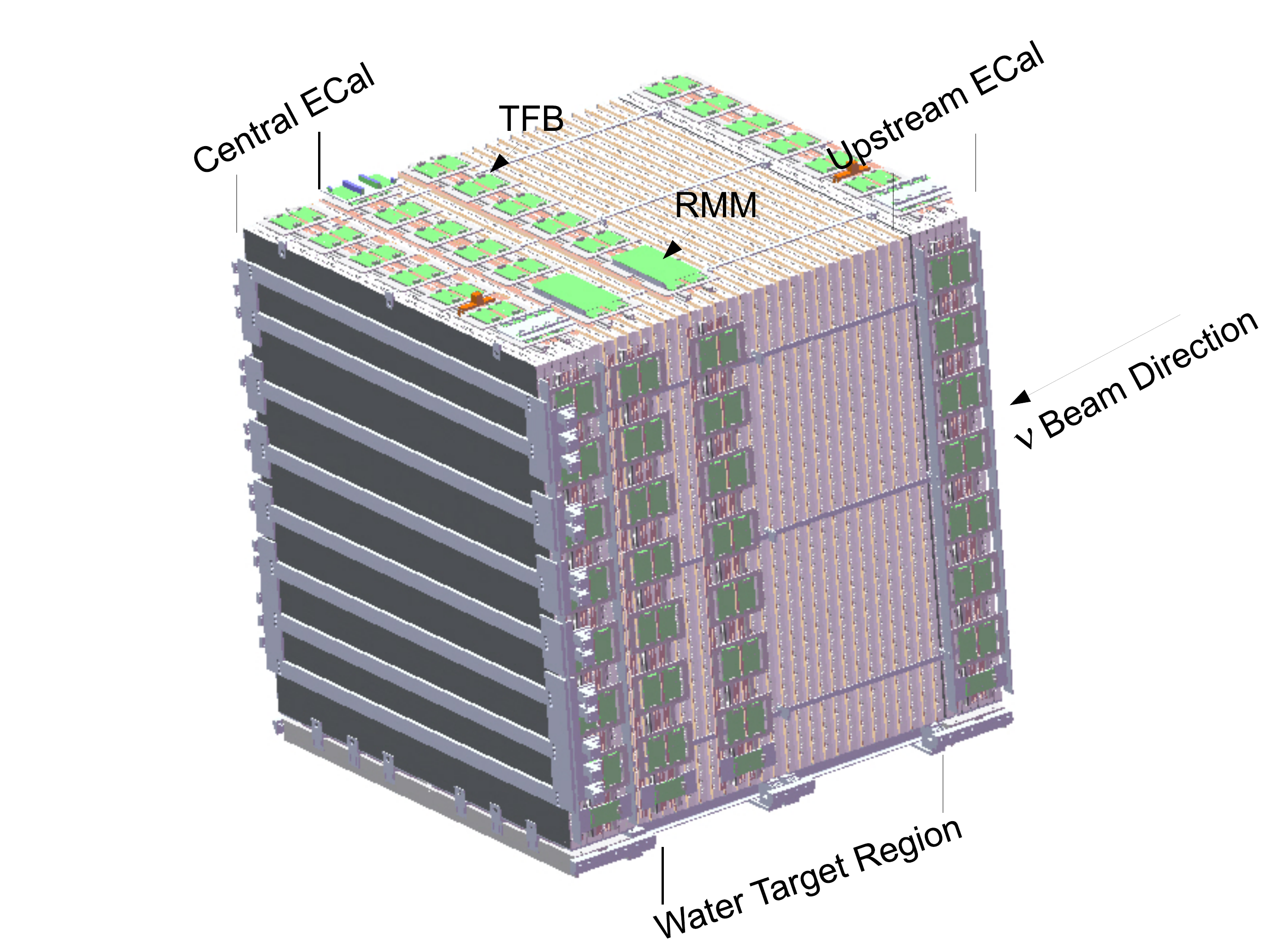}
\caption{3D drawing of the roughly 2.5~m cube \pod{} outside of the basket. Downstream face of detector shown. See Section~\ref{sec_tfb} for a description of the TFB and RMM electronics.}
\label{fig:3dp0d}
\end{figure}

\begin{figure}
  \begin{center}
    \includegraphics[width=0.5\columnwidth]{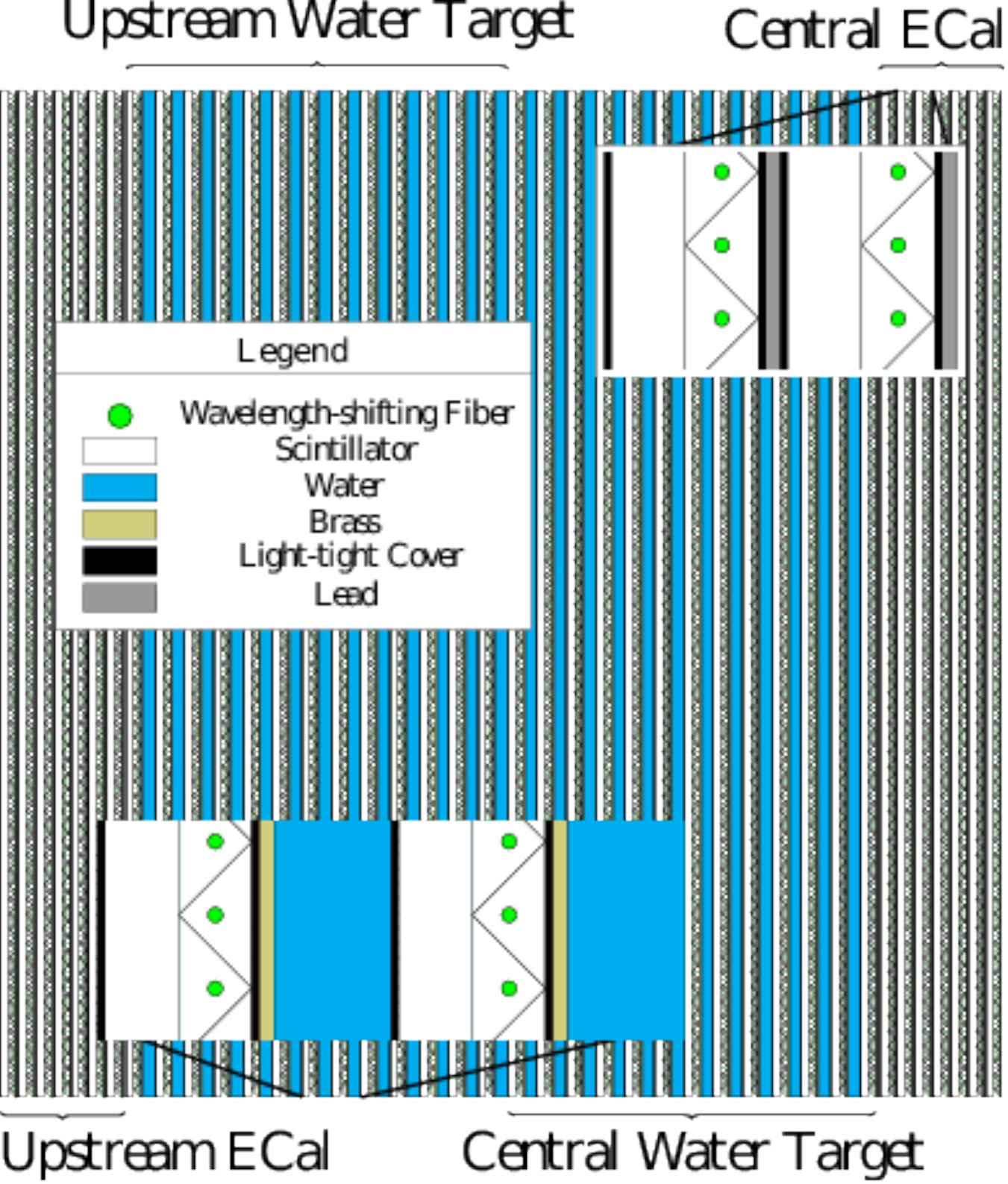}
  \end{center}
  \caption{\label{fg:p0d-schematic} A schematic of the four \pod{}
    \spodule{}s as installed in the detector. Beam direction: left to right.}
\end{figure}

The main features of the \pod{} design are shown in Fig.~\ref{fig:3dp0d}.
The electronics supports and detector mounting system are visible surrounding the active regions of the detector. In addition the regions of the detector are also labeled. Figure~\ref{fg:p0d-schematic} shows a schematic of the active regions of the \pod{}.
The central region, composed of the "upstream water target" and "central water target," is made from alternating scintillator planes, water bags, and brass sheets. The front and rear sections, the
``upstream ECal'' and ``central ECal'' respectively, use alternating scintillator
planes and lead sheets.
This layout provides effective containment
of electromagnetic showers and a veto region before
and after the water target region to provide rejection
of particle interactions that enter from outside the \pod{}.

\begin{figure}
  \begin{center}
    \includegraphics[width=0.5\columnwidth]{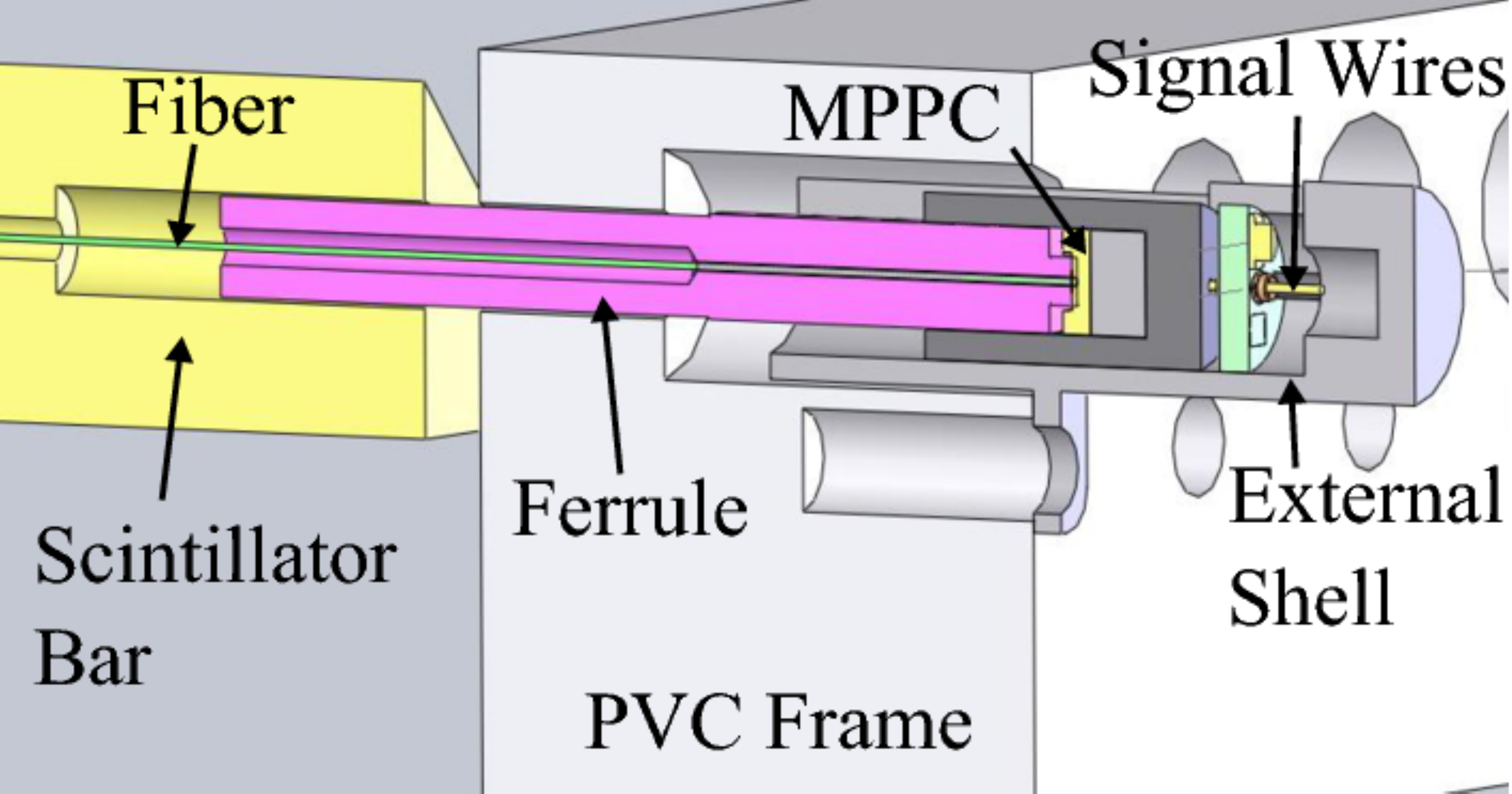}
  \end{center}
  \caption{\label{fig:p0d-mppc} A close-up view of the edge of a \pod{}ule showing how the WLS fibers exit the scintillator bars and couple to the MPPCs.  The optical connectors will be described on more detail in Section~\ref{sec:optical_connectors}.}
\end{figure}

There are a total of 40 scintillator modules in the \pod{}. Each
\pod{} module, or \pod{}ule, has two perpendicular arrays of triangular 
scintillator bars, forming a plane. There are 134 horizontal bars (2133 mm
long) and 126 vertical bars (2272 mm long) in each \pod{}ule.
Each bar has a single coaxial hole through which is threaded a wavelength-shifting  (WLS) fiber. Each fiber 
has a mirrored coating applied on one end while the other end is optically coupled to 
a Hamamatsu multi-pixel photon counter (MPPC)~\cite{Yokoyama:2010qa} for readout, as shown in Figure~\ref{fig:p0d-mppc}. Each photodetector
is read out with Trip-t Front-end electronics (Section~\ref{sec_tfb}). There are
a total of 10,400 channels for the entire \pod{}.

\begin{figure}[htbp]
\centering
\hspace{-1cm}\includegraphics[trim =0mm 0mm 0mm 0mm, clip,width=8cm]{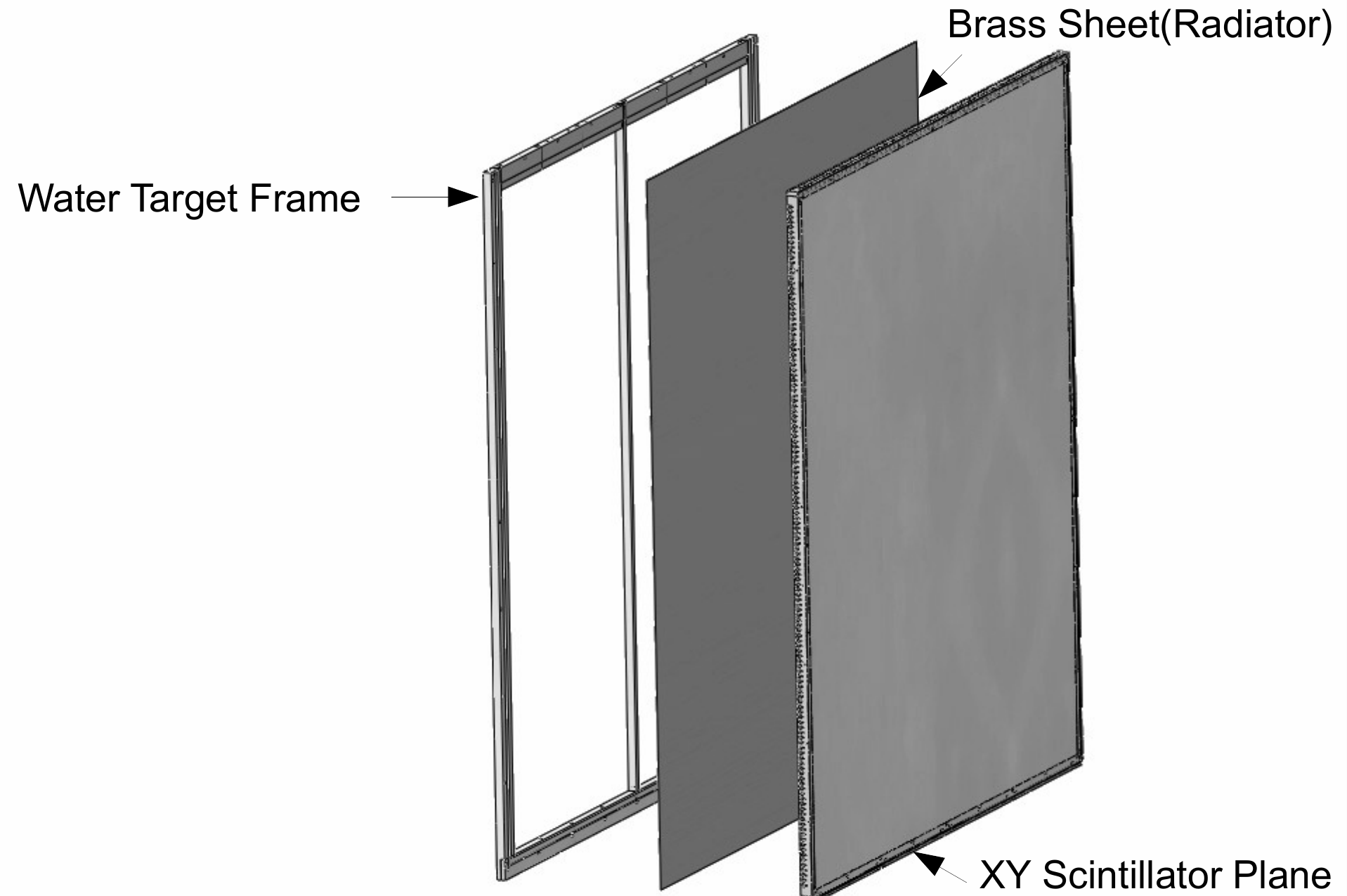}
\caption{Expanded view of water target \pod{}ule, brass radiator and water bladder containment frame.}
\label{fig:wt_p0dule}
\end{figure}

The \pod{}ules were assembled into four units called \spodule{}s. The two ECal \spodule{}s each consist of a sandwich of
seven \pod{}ules alternating with seven stainless steel-clad lead
sheets (4.5 mm thick). The water target is formed from two units, the upstream and central water target \spodule{}s.
The upstream (central) water target \spodule{}
comprises 13 \pod{}ules alternating with 13
(12) water bag layers (each of which is 28 mm thick), and 13 (12) brass sheets (1.28 mm thick), as shown in Fig.~\ref{fig:wt_p0dule}. The dimensions of the entire \pod{} active target are 2103 mm $\times$ 2239 mm $\times$ 2400 mm (width $\times$ height $\times$ length) and the 
mass of the detector with and without water is
15,800 kg and 12,900 kg respectively.  The \pod{} is housed inside a detector basket structure that supports the central off-axis detectors inside the magnet.


The remainder of this paper describes in detail the design, fabrication, and performance of the \pod{}. The production of the scintillator bars and their assembly into planks and \podule{}s will be presented followed by a description of how the individual \podule{}s were combined into the four \spodule{}s, and are read out using photosensors. The detector component performance, starting with scans of the \podule{}s using a radioactive source, dark noise measurements, and tests with the light injection system, is presented. The paper concludes with a description of the calibration and performance of the full detector.




\section{Design and Construction of the P\O{}Dule}




The \podule{} is the basic structural element of the
\pod{} active region, and is constructed of
scintillator bars sandwiched between sheets of high-density polyethelene (HDPE, thickness 6.4 mm).  The entire structure
is surrounded by PVC frames that support the \podule{} as well as providing
mounts for the required services such as the MPPC light sensors, and
the light injection system.

The polystyrene triangular scintillating bars that make up the P\O{}Dules were fabricated by co-extruding polystyrene with a reflective layer of TiO$_2$ and a
central hole for the WLS fiber.  The light seal for the tracking plane is
maintained by light manifolds that collect the WLS fibers into optical
connectors. These manifolds also provide access to the fibers for the light injection
system.  Because of the large number of scintillating bars  and
the available space limitations, it was impractical to route the fibers outside the magnetic volume therefore the Hamamatsu MPPC
photosensors, which are immune to the magnetic field, were attached directly
to each WLS fiber just outside the PVC \pod{}ule frame, as shown in Figure~\ref{fig:p0d-mppc}.


\subsection{Design of the P\O{}Dule}

The \podule{} was designed to both provide the active tracking region and to serve as a structural element.
  This was achieved using a laminated
structure of crossed scintillator bars between polystyrene skins.
The final \podule{} has been shown to have a rigidity similar to a solid
mass of polystyrene of similar thickness.  
The edge of the central scintillator and skin structure of the \podule{} is surrounded by a machined PVC frame.  
Each \podule{} is instrumented on one side (both $y$ and $x$ layers) 
with MPPCs and on the other a UV LED light injection system. 
The bottom PVC frame supports the weight of the \podule{} within the ND280
detector basket.  The frames also provide the fixed points needed to assemble
the \podule{} into the four \spodule{}s via two precision holes
located in the four corners of each \podule{} as well as a
set of seven holes spaced along each side through which tensioning rods
were passed.

The \podule{}s, after installation into the finished \pod{}, are oriented
such that the most upstream layer of scintillator has the bars
oriented approximately along the vertical axis while the downstream layer has its
bars oriented along the horizontal axis. This arrangement results in a local
coordinate system defined such that the $x$, $y$ and $z$ axes are approximately
congruent with the global coordinate system where $x$ is horizontal, $y$ is
vertical, and $z$ points downstream toward Super-Kamiokande.
The external dimensions of the \podule{} are 2212\unit{\milli\meter} ($x$) by
2348\unit{\milli\meter} ($y$) by 38.75\unit{\milli\meter} ($z$).

To facilitate  assembly of the \podule{} (described in
Section~\ref{sec:p0dule-assembly}), all of the components were prefabricated
with holes that allowed alignment during assembly.  The
assembly tolerance was less than 0.5\unit{\milli\meter} on all internal
dimensions, and less than 1\unit{\milli\meter} on the thickness. The
relative dimensions of the \podule{}s were maintained using precisely located 
holes in the \podule{} assembly table.

\subsection{Assembly of P\O{}Dule Components}

The construction and assembly of the P\O{}Dule components was distributed across several institutions. This allowed a supply chain that could produce the required
components in parallel and optimized use of facilities, local
expertise and available personnel.

\subsubsection{P\O{}Dule Scintillator Preparation}

The polystyrene scintillator for the P\O{}Dule was manufactured in the
extrusion facility at Fermi National Accelerator
Laboratory \cite{Dyshkant:2006us} using an extrusion die
originally developed for the inner detector of the MINER$\nu$A
experiment \cite{Drakoulakos:2004gn, McFarland:2006pz}.  The blue-light 
emitting scintillator base material was Dow Styron 663 W doped
with $1\%$ PPO and $0.03\%$ POPOP to shift the UV scintillation light
emitted by the polystyrene into the blue.  The bars are triangular in
cross section with nominal dimensions of $17\pm 0.5$~mm height and
$33\pm 0.5$~mm width.  Each bar also had a nominal $2.6$~mm diameter hole
centered at $8.5\pm0.25$~mm above the widest part of the
triangle for fiber insertion.  To reflect the produced light and therefore maximize the
possibility of capture by the wavelength shifting fiber in the center
hole, a thin, $0.25$~mm on average, layer of polystyrene mixed with $25\%$
TiO$_2$ was coextruded on the outside of the bar, and both ends of
the scintillator bar were painted with white EJ-510 TiO$_2$ Eljen
paint.

During production, physical characteristics of the scintillator were
monitored by taking frequent samples and measuring their outer
dimensions, the location and dimensions of the center hole, and the
thickness and coverage of the coextrusion.  At approximately twenty
different equally spaced times during production, samples were also
taken and used to characterize light output using a radioactive source
counting setup with a reference piece of scintillator from the MINER$\nu$A
production.  Physical dimensions were held well within the tolerance,
and no evidence was observed for variation in light output beyond the
uncertainties in the monitoring measurement, roughly $5\%$.

\subsubsection{P\O{}Dule Plank Assembly}

The extruded scintillator bars were bundled into manage\-able sized ``planks'' to be used in the assembly of the
full-sized P\O{}Dules. There were two sizes of planks for each of the bar
lengths and a special jig was constructed for each of the four plank
types. 

Each of the triangular scintillating bars was prepared for the plank
assembly and subjected to quality assurance (QA) procedures prior to
assembly into a plank. As each bar was unpacked it was inspected for
signs of visible damage, such as nicks or cuts in the TiO$_2$
coating and any damaged bars were removed from plank production. Once a
bar passed the visual inspection it was cut to length 
using a jig to
ensure proper length. A mounted pneumatic drill was used to bore out the ends of the holes running down the
center of each bar. A long stiff wire was passed through each bar to
ensure that no debris was lodged in the hole that would prevent insertion of the WLS fiber. An additional check was made
to ensure that the hole for the fiber was centered on the end of the triangular bar. 

Four separate jigs were set up on two optical tables for the gluing of
the bars into planks. The short bars were made into two types of
planks, one type containing 16 bars and one type containing 17
bars. The long bars were made into 15 bar and 17 bar
planks.  Prior to application of the epoxy to the bars, the
bars were placed in the jig and a heavy straight-edge was used to ensure that the
thickness of bars were within the 0.25~mm tolerance of the
nominal 17.25~mm plank thickness. A log document, or traveler,
was kept with each plank during the entire assembly process. The
traveler contained details such as the plank serial number, the
identification number of all bars contained in the plank, and any
measurements made on the plank during assembly and quality assurance. 

\begin{figure}[htbp]
\centering
\includegraphics[width=2in]{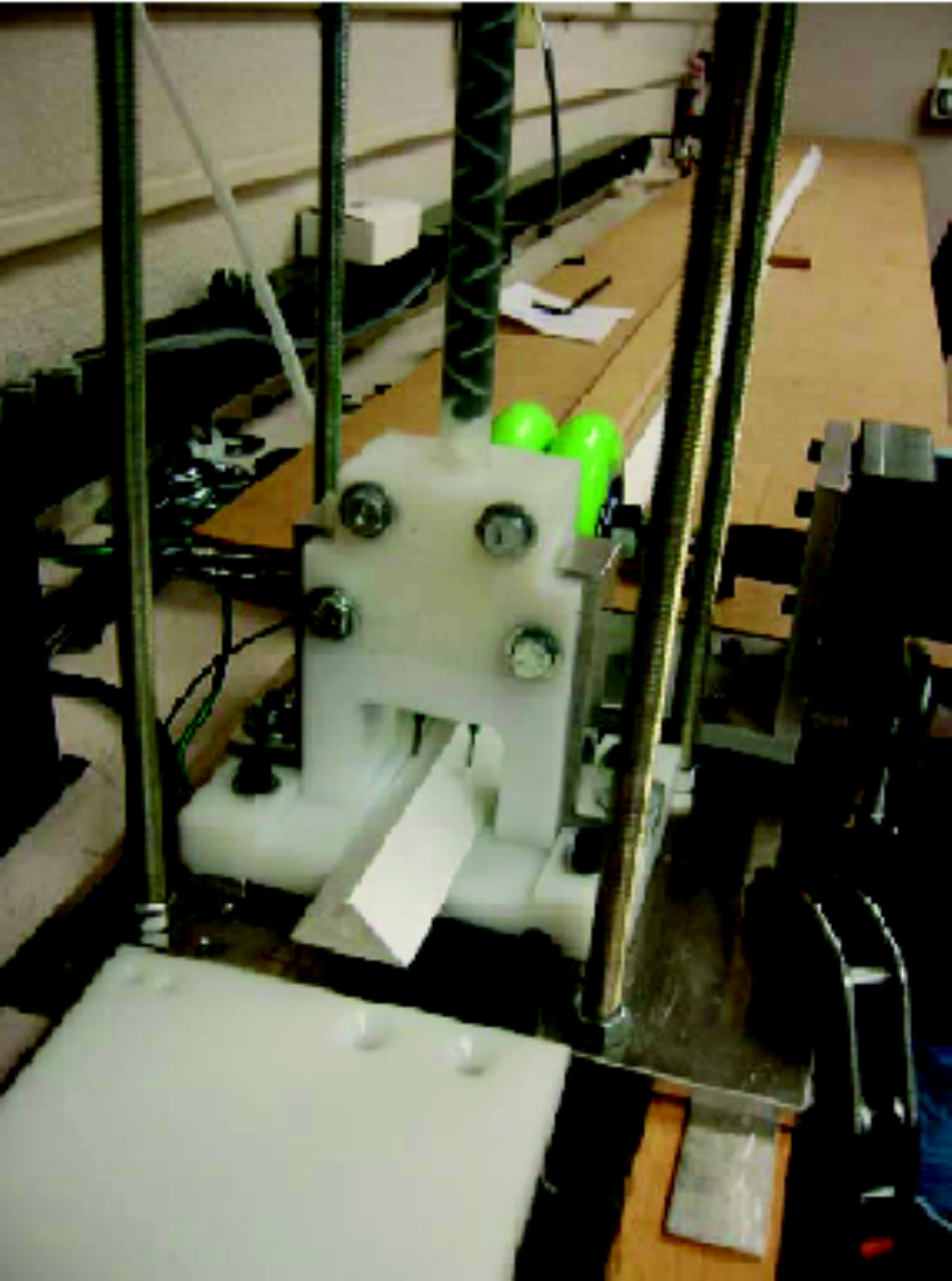}
\caption{Automated glue machine used to apply epoxy to the triangular
bars before they were placed into the plank gluing jig.}
\label{fig:glue}
\end{figure}

Once the bars had been test fitted into the plank gluing jig, they
were removed and epoxy was applied to each using an automated gluing
machine (Fig.~\ref{fig:glue}) that mixed the two epoxy parts and
applied a steady stream of glue to two sides of the bar. The
epoxied bars were placed back into the gluing jig and a vacuum sealed
frame was used to apply pressure to the plank for about 2 hours while the
epoxy set. A final QA inspection was made to ensure that the planks
were within the thickness tolerance. 

\subsubsection{P\O{}Dule WLS Fiber Preparation}

The wavelength shifting fibers that are inserted into the holes in the scintillator bars are Kuraray multi-clad, S-35, J-type, doped with Y-11 (175 ppm)
with a diameter specification of $1.00^{+0.02}_{-0.03}$ mm.  The fibers were placed into the holes in the scintillator, but the hole was not filled with optical glue after the fiber was put into place.  Studies done for the Miner$\nu$a experiment indicated that the light yield for glued fibers was approximately 2 times greater than the light yield for unglued fibers~\cite{PlaDalmau:2005}, but unglued fibers are considerably easier to install.  The same study also showed that the light yield did not strongly depend on the fiber-to-hole diameter ratio over the range of 0.3 to 0.9~\cite{PlaDalmau:2005}, so the decision to use a 1.0 mm fiber in a 2.6 mm hole does not have a large impact on the light collection.

The WLS fiber  was delivered in unspooled
``canes'' pre-cut to a rough length 67 mm longer than the bar length in
order to avoid memory effects of spooled fiber.

The first step in processing the delivered fiber was to mirror one
end.  This work was performed in the Thin Film Coatings facility in
Lab 7 at Fermi National Accelerator Laboratory.  One end of the fiber was first
``ice polished'' with the ice providing mechanical support for a
group of approximately 800 fibers polished with a diamond
polisher in a single batch.  The polished end was then coated with
aluminum using a sputtering vacuum deposition process.  After
completion of the mirroring, each fiber was coated with a thin layer of
epoxy to protect the mirror.   

The reflectivity of three
fibers from each batch of 800 was determined by measuring the light output of a fiber
with the mirror end placed into a piece of scintillator with an attached radioactive
source, and then remeasuring the light output after cutting off the
mirror with a $45^\circ$ cleave and painting the cleaved end with black
paint.  For individual fibers,  an average reflectivity of
$86\%$ was measured, with a root mean square of the ensemble of measurements of
$6\%$.
 
After mirroring, the opposite fiber end was glued into one end of a ferrule (see Section~\ref{sec:optical_connectors} and Figure~\ref{fig:optical-connector}) that was
injection molded from Vectra A430, a Teflon-filled liquid-crystal
polymer.  The length between the ferrule end and the fiber mirror end
was required to be kept at $1$ mm tolerance over the several meter
length of the fiber.  The ferrule was designed to mount into a housing that
contained the MPPC and kept the fiber end in contact with the pliable optical
layer covering the MPPC.  Because the Teflon in the Vectra
plastic clogged the diamonds in the fiber polisher, the fiber was glued 
into the ferrule a few mm proud of the end of the ferrule,  a
1 mm thick layer of optical epoxy was deposited for mechanical support, and then
the epoxy with the embedded fiber was polished.  The resulting finish was
inspected with a microscope to establish
when diamond wear  adversely affected the finish.  Typically, 1500 to 2000 fibers could be polished
with a single diamond.

\subsubsection{MPPC Acceptance Testing}

The MPPCs used by the \pod{}, as well as by the ECAL, SMRD, and INGRID detectors, are solid-state photosensors
manufactured by Hamamatsu Photonics. The active sensor for the MPPC  is a
1.3 mm$\times$1.3 mm array of 50-micron pixels, totaling 667~pixels (including
a small inactive area for electrical contact).  Each pixel operates in
Geiger mode, producing a well-defined pulse when a photoelectron generates a cascade. The MPPC
output is the sum of all pixel outputs, the pixels having very similar responses.  As a result, the output
spectrum from the MPPC shows well-separated peaks corresponding to
the number of pixels fired, which at low light levels is a good measure
of the number of photoelectrons.

A photosensor quality-control (QC) testing program was performed on all
10,400~MPPCs installed in the \pod{}, as well as 1,100 spares.
The goals of the testing were to measure the operational characteristics
of the MPPCs, to verify that their performance was acceptable,
and to set the initial operating voltages to be used in the \pod{}.  In particular, the gain of each device
was measured as a function of bias voltage. A linear fit was used to extract the
breakdown voltage -- the minimum bias voltage to produce a Geiger cascade.
The MPPC characteristics were found to be largely consistent from device to device as
a function of overvoltage (the bias voltage above the breakdown voltage),
but there is a significant variation in breakdown voltages from device to
device, particularly between devices originating in different batches. All tests
were done at a controlled temperature of 20$^\circ$C, controlled to better
than 0.2$^\circ$C, as the MPPC breakdown voltage depends on the
operating temperature.

The testing protocol required the measurement of a number of different MPPC operating
parameters.  First, a scan of gain vs bias voltage was performed to measure the
breakdown voltage and to establish the operating voltage.  
Measurements were made over a 2 V wide overvoltage
range around the predicted gain range of $5\times10^5$ and $7.5\times10^5$. 
Dark noise rates and relative detection efficiencies
at these gains were also measured.

Physics signals in the \pod{} range from a few to hundreds of photoelectrons. Eight light
levels across this range were used to characterize the photosensors, with
measurements made at four different bias voltages for each light level.


Production testing of the photosensors began in late September 2008 and was
completed by January 2009. Photosensors passing all QC tests were shipped for installation into the completed \podule{}s. Only
230 out of 11,500 photosensors were rejected by the quality control
procedure, despite stringent acceptance criteria.  Of those rejected,
74 were broken during assembly, 76 were rejected due to dirt on the
photosensor surface. Only 80 were rejected for abnormal behavior in
the photosensor testing data. 

A dedicated NIM paper describing the photosensor quality-control
testing and characterization procedure in more detail can be found at~\cite{mppc_characterization}.  A paper describing the \pod{} MPPC testing in greater detail is under preparation.

\subsubsection{Optical Connectors}
\label{sec:optical_connectors}

\begin{figure}[htb]
 \centering\includegraphics[width=.45\textwidth]{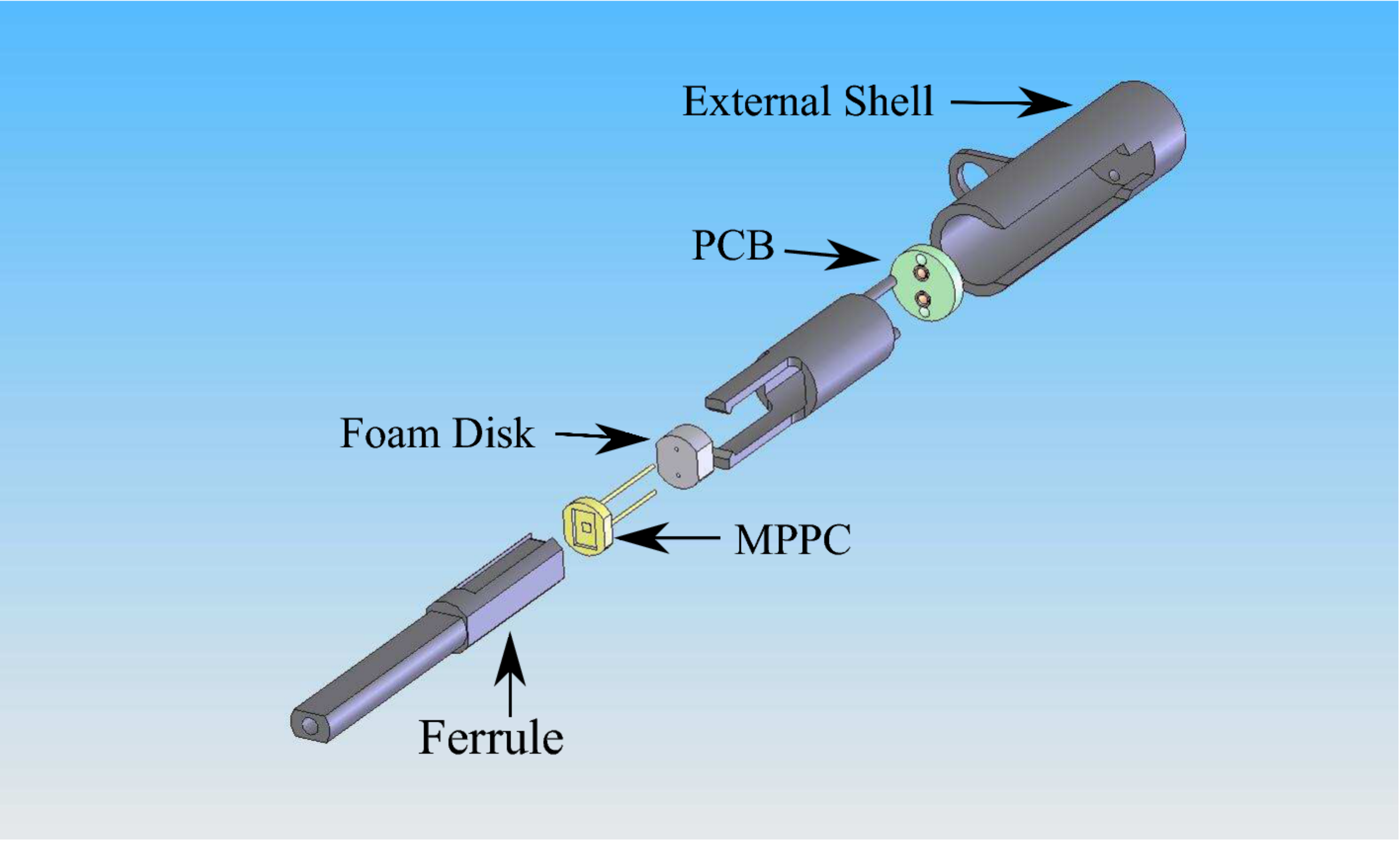}
 \centering\includegraphics[width=.45\textwidth]{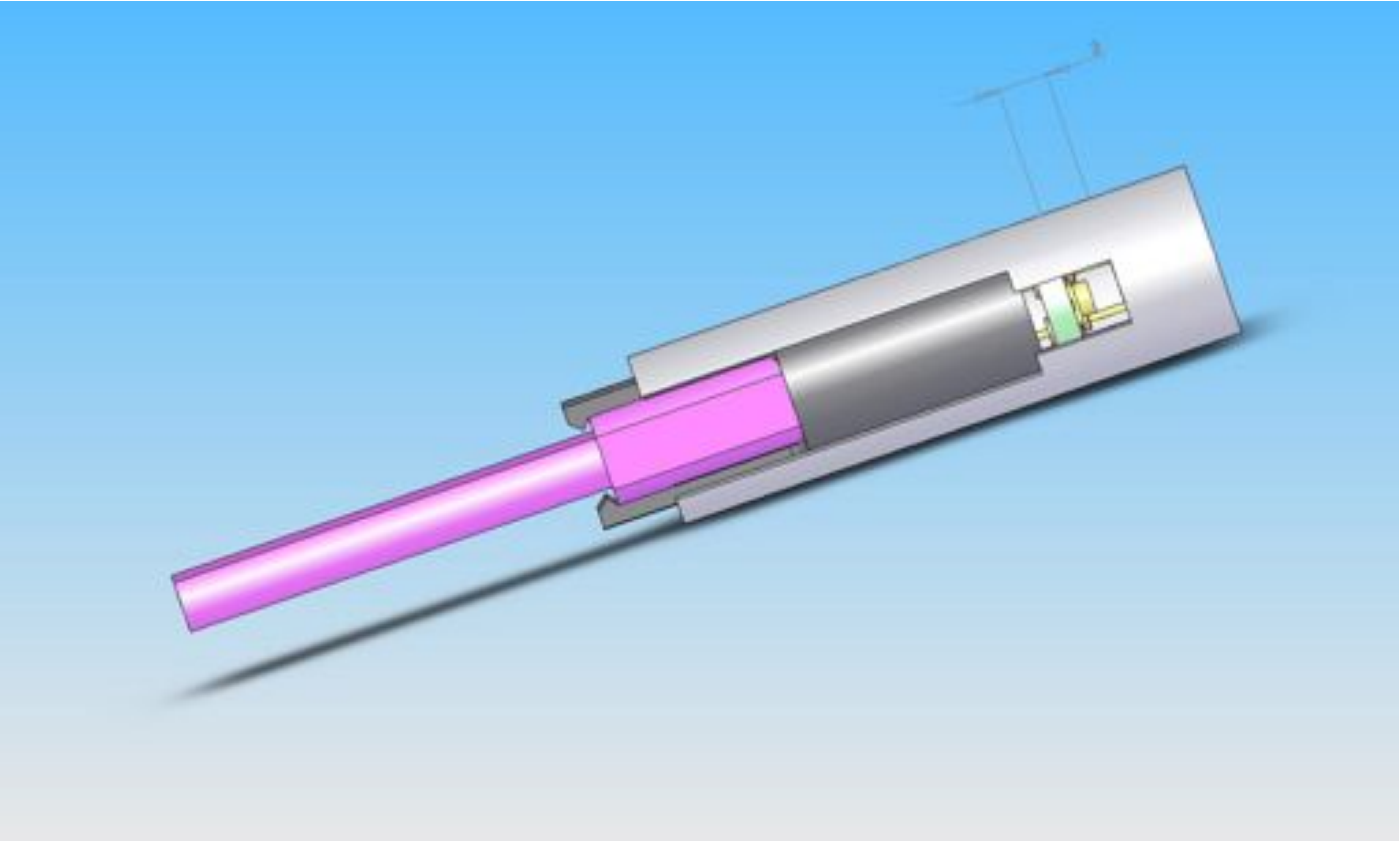}
  \caption{Optical connector: exploded (left) and assembled (right).}
  \label{fig:optical-connector}
\end{figure}

Custom optical connectors were fabricated to provide optical fiber alignment to the MPPC active area and to reduce the signal rate due to light contamination from external sources to acceptable levels to well below  the intrinsic dark noise of the MPPCs.  The connector system, shown in Fig.~\ref{fig:optical-connector} consists of three injection-molded components:  a fiber-alignment tube or ``ferrule'', a housing  that holds the MPPC and provides alignment to the fiber within 150~$\mu$m, and an external shell to provide mechanical protection and lock the connector on the ferrule.  The material for all three molded components is Vectra\textregistered A130, a 30\% glass fiber loaded liquid-crystal polymer with very low shrinkage, excellent dimensional stability, and very good mechanical properties.

The MPPC is held in place against the fiber by a 3~mm thick closed-cell polyethylene foam disk, acting as a spring.  To ensure good optical contact between fiber and MPPC, the fiber is allowed to protrude from the ferrule end by 0.5~mm after polishing, so the entire compression force from the foam spring is applied between the fiber end and MPPC face. Electrical connection between the MPPC and the front-end electronics is provided via a small circular printed circuit board (PCB), with spring-loaded pin sockets making contact to the MPPC leads and a Hirose Electric Co. micro-coax connector.

Connectors of this type are used in ND280 for the P0D, ECAL and INGRID sub-detectors.  Approximately 40,000 total connectors were produced (including spares).

\subsection{Assembly of the P\O{}Dule}
\label{sec:p0dule-assembly}

The \pod{}ule assembly was one of the main fabrication stages to produce  the forty
\pod{}ules (plus 10\% spares). This step involved gluing the main mechanical
components: the
pre-glued scintillators planks, the four outer PVC
frames, and the two outer HDPE plastic skins. A key requirement for this process was to
keep within the tolerances of the thickness of the \pod{}ule (28 mm) and to have the
alignment of the scintillator bars match the wavelength shifting fiber
holes in the scintillators with the holes in the PVC frames.

\pod{}ule construction was performed on a specially designed gluing table. This flat
table ($\sim$ 2.5 $\times$ 2.5 m) had precision alignment holes to keep the
outer four PVC frames in the same location for all the \pod{}ules. The
first assembly step was the placement of the HDPE plastic sheet or skin
on the gluing table, followed by painting HYSOL epoxy glue with paint roller
brushes over the entire top side of the sheet. Next, the four outer PVC
frames were placed on the edges of the glued sheet, and the
position of the PVC frames were fixed by steel pins pressed into the holes
in the table. Next, the scintillator $x$-layer planks were lowered onto the bottom
HDPE sheet. They were aligned with respect to
the PVC frame holes, then steel alignment pins were pressed through the frame
holes and into the holes in the scintillator. After all the $x$-planks were
pinned into place, epoxy glue was painted over the upper surface of the
$x$-plank scintillator. Then the $y$-planks were lowered onto
the epoxied $x$-planks, and again the $y$-plank
positions were aligned with steel alignment pins. Next
 the upper HDPE outer sheet or ``skin'' was lowered on to the glue-painted y-plank
surface.

After all the gluing was completed, a vacuum sheet was lowered onto the
\pod{}ule and vacuum was applied so the \pod{}ule was uniformly compressed on all
sides with about 0.5 atmosphere of pressure.  The \pod{}ule was left for about 12
hours to cure overnight. After the epoxy had cured, the
vacuum sheet and the pins were removed and the \pod{}ule was loaded with 260
WLS fibers that were each attached to the snap-on optical
connector containing an MPPC. Finally the 260 MPPCs were connected via
mini-coax cables to the scanner readout  electronic boards. The
\pod{}ule was scanned overnight with a $^{60}$Co source and
read out as described in Section~\ref{sec_scanner}. 

During the peak of the construction phase, five \pod{}ules were produced per week.
After the \pod{}ules were constructed
and successfully scanned, they were stacked together vertically
into the four Super-\pod{}ules.

\subsection{The P\O{}Dule Light Injection System}

The purpose of the \pod{} Light Injection System (LIS) is to provide monitoring for all 10,400 channels in the \pod{}. 
The LIS is capable of exposing the MPPC photosensors to light intensities covering a range of more than two orders of magnitude with flash-to-flash intensity stability of less than 2\% ~\cite{minosLI}. 
This allows for monitoring of the photosensor response over the full range of energy deposition expected for neutrino interactions in the \pod{}. 

\begin{figure}[h]
\centering
\includegraphics[width=0.3\textwidth]{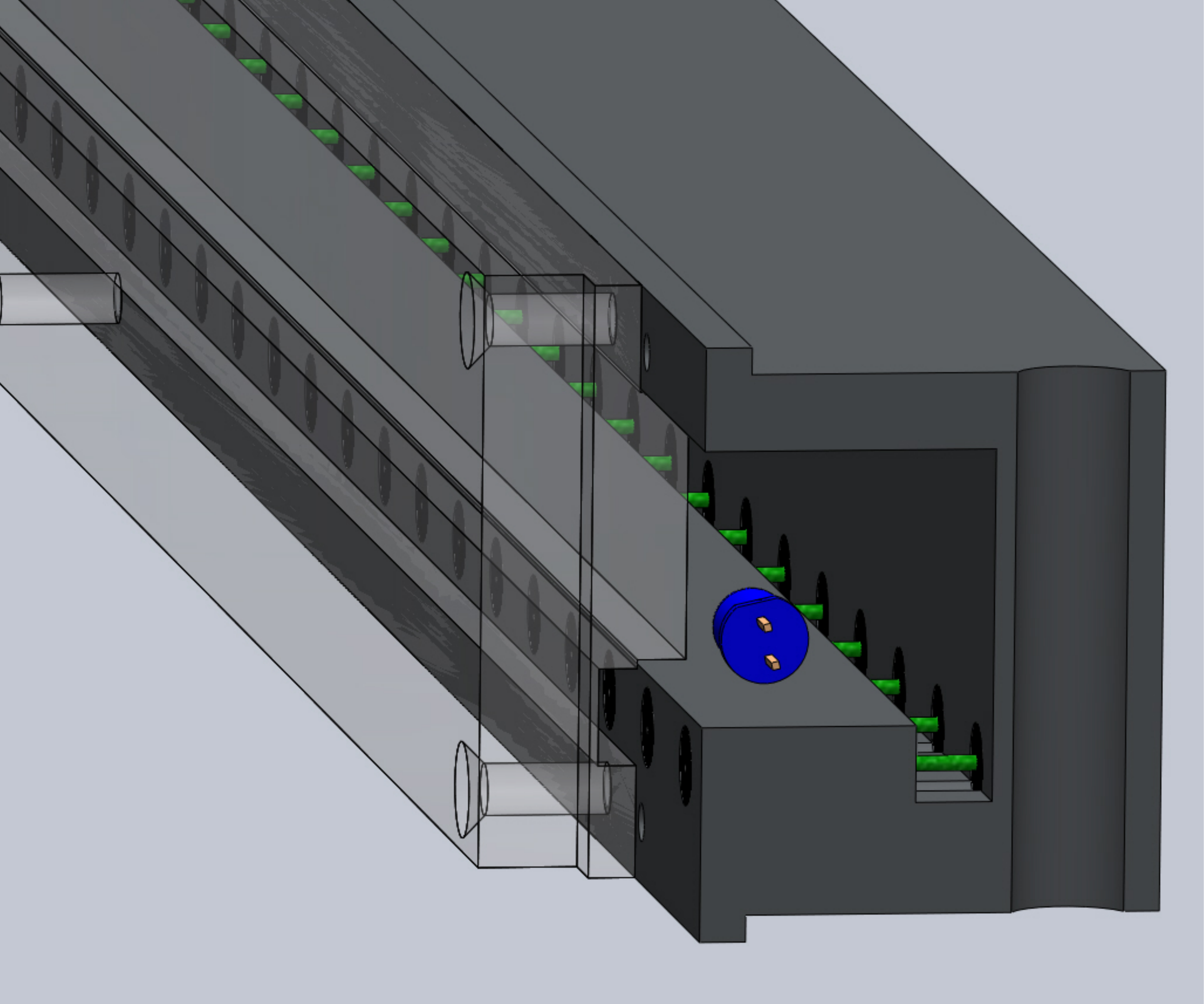}
\caption{Cross-section view of the LIS cavity. Not shown are transparent strips with an opaque band to shadow nearby fibers.} \label{p0duleli_figure1}
\end{figure}

The main design challenge of the LIS was the geometrical constraint that it be embedded within the 3 cm $\times$ 4 cm $\times$ 220 cm \pod{}ule layer support frame for each of the 80 scintillator planes. 
Each \pod{}ule support frame has a cavity that allows 5 mm segments of WLS fibers to be exposed to light produced by a pair of LEDs (Fig.~\ref{p0duleli_figure1}). 
The LIS uses 80 pairs of fast-pulsed 400 nm UV LEDs as light sources;  each pair illuminates a cavity in a single \pod{}ule support frame. 
LEDs exhibit minimal pulse-to-pulse fluctuations in intensity, so the temporal response of an MPPC photosensor is dominated by photoelectron statistics. 

The LIS provides the capacity to vary the LED light intensity over the required dynamic range through control of both the height and the width of the pulses, by varying the current pulse applied to the LED. 
The LEDs are driven by electronics originally designed for the MINOS experiment LIS system~\cite{minosLI}. 
The LIS electronics consists of four pulser boxes, a control box, a distribution box and a power supply. 
The pulser boxes are mounted on the bottom and the north sides on the \pod{} in close proximity to the LEDs. 
The control box is situated under the \pod{} , mounted on the detector basket.
The distribution box and the power supply are located in a rack  about 10 meters from the \pod{}.
Each pulser box contains two LED driver boards, a controller board, and an LVDS to TTL converter board. 
Each driver board has ten channels that can be programmed to pulse simultaneously in group of of 5, 10, or 20 channels.  In contrast to the MINOS light injection system, each channel is connected to a pair of LEDs by a 60 cm long shielded cable. 
Communication between a PIC16F877 microprocessor located on the controller board and
a control computer is handled via ASCII commands over a serial RS232 link. 
Signals from the control computer are carried over an Ethernet link and converted for the pulser boxes by an Ethernet-RS232 converter.

During normal operation the \pod{} LIS control computer instructs the system to pulse a specific combination of LEDs at a specific height, width, and frequency to monitor the temporal behavior of the \pod{} and associated readout. Dedicated LIS runs are taken periodically to allow more precise measurements of properties, such as the relative timing between the \pod{} channels.

\section{Water Targets}




\subsection{Water Target Design}

The water target modules were designed to provide layers of water approximately 3 cm thick in the beam direction, with transverse dimensions of approximately 2 m $\times$ 2 m. In order to provide such large but thin layers without excessive container mass, high-density polyethylene (HDPE) bladders were fabricated using materials and techniques derived from experience with the Pierre Auger Observatory Water Cherenkov detector liners~\cite{Auger}. Adjacent scintillator bar layers provided the structural support required, with water pressure transmitted through successive \podule{}s to end plates 
attached to the ND280 basket frame. 

Since the water bladders are not rigid, it is necessary to protect against excessive deflection of adjacent scintillator bar layers due to unbalanced pressure, such as the case where one water layer is emptied while the others remain  full. A vertical HDPE center strut to limit such deflections was required, so each water layer consists of two side-by-side bladders, each approximately 1 m wide by 2 m high by 3 cm thick.

The bladders were heat-welded and sealed into HDPE frames that provide ports for fill, drain, and sensor tubes. The frames also accommodate the mounting and support hardware used to assemble the \spodule{}s. The frames were edged with silicone sponge gasket material (Stockwell Elastomerics R-10470), such that each bladder was effectively encased within a waterproof seal when compressed between adjacent \podule{}s. This provides a second level of water containment to protect against leaks. The water bladder frames were equipped with drain ports to direct any water captured within the gaskets into two drip pans mounted below the \pod{} and basket, one on either side of the basket centerline. 
Each target layer was filled with water and left overnight prior to acceptance and shipment for integration with scintillator layers and assembly into \spodule{}s.  
The upstream target \spodule{} had 13 water layers (26 bladders) and the central target \spodule{} had 12 layers (24 bladders), for a total of 25 water target layers and 50 bladders. After integration, water fill/drain testing was repeated and repeated a third time when the \spodule{}s arrived at J-PARC.

In order to meet the physics analysis requirements, a water pump and monitoring system was designed to allow individual bladders to be filled, drained, and also provide water depth data. A pump rack with 50 self-priming bellows pumps (Gorman-Rupp Industries (GRI) 16001-005 F-009 T-007), one for each bladder, was designed with valves arranged to allow each pump to be used either to fill or drain a single bladder. The pump rack is located outside the magnet. The pumps are connected to fill and drain tubes on the bladders using Polyflo 66P 3/8-inch OD polyethylene tubing. 

Water is supplied from a 3000 liter high-density polyethylene tank (DenHartog model VT0900-46), next to the pump rack, coupled to the pumps through two 20-liter buffer tanks on the rack. The main tank is filled with filtered tap water, with commercial chlorine bleach added as a biocide at an effective concentration of 0.025\% sodium hypochlorite by volume. 

Each bladder is equipped with separate drain and fill tubes made of 3/8-inch schedule 40 PVC pipe, and has two additional PVC pipes containing water monitoring sensors of two kinds: binary level sensors, which simply report their state as wet or dry, and depth (pressure) sensors.

\subsection{Fabrication of Water Targets}

\bigskip

The water target bags are made from a 1 meter wide continuous roll of
seamless polyethylene plastic tubing. The plastic sheet is 0.1 cm thick, so
the combined thickness of 2 walls of the water bag is 0.2 cm. 
The tubing is cut to the length of the \pod{}ule height, and is heat welded to make a bag with a leak proof seal at
the bottom with no seams on the sides of the bag. Two HPDE frames are then
attached, one on the top and one on the bottom section of the bag. The top
frame is slotted to provide entrance holes from the top for the water tubes
and the pressure and level sensor pipes. The top HDPE frame length is held
in position by the HDPE gasket frames on the \pod{}ule. The bottom frame is
threaded to allow for through-screws, protruding from the
bottom HDPE frame of the gasket frame, to pull the bag taut in the
enclosure. This arrangement allows for a leaky bag to be pulled straight from the top of the water target \spodule{} and replaced by lowering a new
bag from the top without removing the \spodule{} out of the basket frame.

\subsection{Instrumentation of the water targets}

In order to properly monitor the water level in each water target layer, sensors are inserted into each water target bladder. Accurate water level readings are needed not only for safety and engineering concerns, but also to determine the water mass inside the fiducial volume of the \pod{} to the desired accuracy of 3\%.

The monitoring system consists of sensors that are inserted into the bladder and operated in the water, an external monitoring sensor for the environment, and a DAQ system. 
The entire system is designed to be independent of the ND280 Global Slow Control system in order to be independent of shutdowns and to provide additional flexibility needed during \pod{} fill or drain operations.

Two sensor pipes are installed in each of the 50 water bladders.  In each bladder, both pipes have a Global Water WL400 water depth sensor at the bottom end of the pipe.   Each pipe also have one Honeywell LLE series Liquid Level Sensor binary wet-dry level sensors placed near its top, for calibration and back-up purposes.  
All new sensor pipes were water tested overnight or longer prior to deployment.
The depth sensors are calibrated by filling test pipes to a series of measured heights, and logging repeated depth readings. These data are then used to fit calibration curves relating depth to current-loop mA. Results are consistent with factory calibration data. The calibration process is checked with a 1-point measurement using an identical test stand after shipment to J-PARC. 

The water sensors required an auxiliary, custom built, connection board to distribute DC power and convert current-loop signals to voltage signals to be read out by the sensor boards. The sensor boards are mounted on the top of the \pod{}, above the water target bladders. This board contained a SenSym ICT series ASDX Pressure Transducer, a Texas Instruments TMP275 Digital Temperature Sensor, 12-bit ADCs to read the sensor outputs, and digital logic to communicate with the rest of the DAQ.  Communication between the DAQ and system hardware components is performed using I\squared{}C. Signals are 
converted to RS-232 serial data for transmission due to the distance between the \pod{} and the DAQ computer. Sensor signals were digitized on the Sensor Boards, which were in turn connected to a custom built, 8 port Multiplexer (Mux) Board using the I\squared{}C bus. A total of four Mux boards are required. The Mux Board converted the I\squared{}C signal to RS232 and also supplied power to the sensors and electronics.  The sensors are controlled and monitored by a graphical interface 
using the Qt 4.0 application and user interface.  
The monitoring program controls the sensor readout, interprets the data, feeds the data to the GSC system, and then stores the data locally.  

An identical WL400 sensor and readout board is installed in the main water storage tank in order to monitor its water level during filling and draining.  

\section{Super-P\O{}Dules}




The full \pod{} detector is constructed of four \spodule{}s
assembled from \podule{}s.    Figure~\ref{fg:p0d-schematic}
shows the arrangement of the four \spodule{}s in the assembled detector as
well as details of the structure within each type of \spodule{}.

\subsection{The Super-P\O{}Dule Design and Tooling}

The four \spodule{}s that make up the \pod{} were designed to simplify the
installation and shipping from the assembly site to the installation
site at J-PARC while allowing for efficient assembly.  This was achieved by
assembling all four modules from a set of standardized components
(\podule{}s, water target modules, brass radiators, and lead radiators)
that could be preassembled prior to the \spodule{} assembly.  The
components were assembled onto their final mounting hardware,  which was held
on custom rolling carts.  The carts were designed to support the \spodule{}s
and allow them to be moved.
After arrival in
Japan, the \spodule{}s were kept on their individual carts which allowed
them to be moved and tested independently.

Table~\ref{tb:sp0dule-mass} shows the mass, dimensions and the
depth in radiation
lengths of each \spodule{}.  During assembly the
component masses were sampled and used to estimate the dry mass of each
\spodule{}.  The mass of the water added to the target \spodule{} is
accounted for separately.  We estimate that the dry mass uncertainty is
approximately 0.8\%.  The length of each \spodule{} along the neutrino beam
axis was measured after assembly and is estimated to have an accuracy of
0.5\unit{\milli\meter}.  The width and height dimensions of the \spodule{} are
perpendicular to the beam axis and have a tolerance of
5\unit{\milli\meter}.  These dimensions include space for the TFB read out
electronics.

\begin{table}
\begin{center}
\begin{tabular}{|l|c|c|c|}
\hline
\spodule{} &Mass  & Dimensions  & Depth \\
&(kg) & (mm$\times$mm$\times$mm)& in R.L.\\
\hline
\bf{Upstream ECal}       & 2900 & 2298$\times$2468$\times$305 & 4.946 \\
\bf{Upstream Water} &  & 2298$\times$2468$\times$888 &  \\
\bf{Target}: & & & \\
Empty & 3600 & & 1.370\\
Filled & 5100 & & 2.379 \\
\bf{Central Water}  & & 2298$\times$2468$\times$854 & \\
\bf{Target}: & & & \\
Empty & 3500 & & 1.356 \\
Filled & 4900 & & 2.287 \\
\bf{Central ECal}       & 2900 & 2298$\times$2468$\times$304 & 4.946 \\
\hline
\end{tabular}
\end{center}
\caption{The mass, dimensions, and depth in radiation lengths for
  each \spodule{}.}
\label{tb:sp0dule-mass}
\end{table}

\subsection{ECal Super-P\O{}Dule Assembly}

\bigskip

\begin{figure}[h]
\centering
\includegraphics[width=80mm]{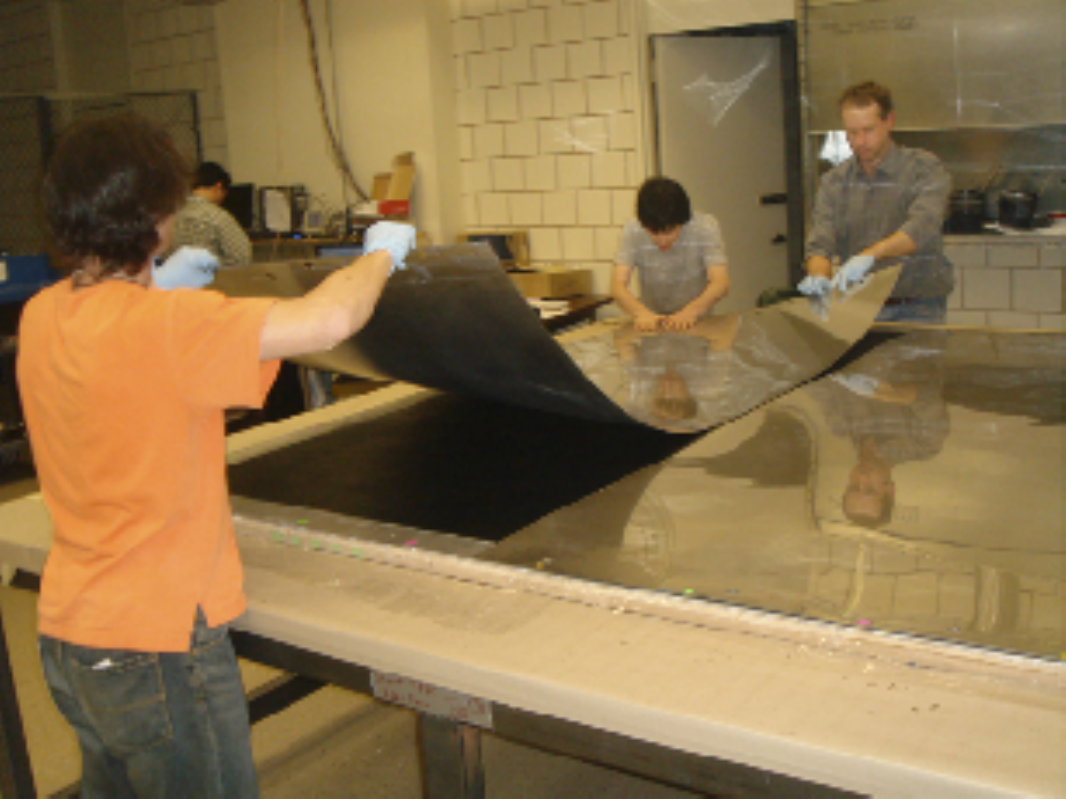}
\caption{A stainless steel plate being placed onto a lead panel} 
\label{fig-lead-plate}
\end{figure}

An ECal \spodule{} is assembled from seven \pod{}ules interleaved with seven layers of lead plates.
The assembly began with the construction of the lead panels. The first step
was the preparation of a 0.05 cm thick, or 0.03 radiation lengths (R.L.), stainless steel (S.S.) sheet
which was mounted on a flat assembly table. A thin aluminum frame was then epoxied and  screwed on all four sides of the sheet. Epoxy was applied to the
surface of the S.S. sheet and 25 3.45 mm thick (0.67 R.L.) lead strips were
gently positioned side-by-side onto the epoxied surface. The lead plates were precut to size to minimize any gaps. The lead plates were then painted with epoxy and two stainless steel half-width sheets were placed onto the pre-epoxied lead plates, 
creating a panel that contained lead plates sandwiched between a pair of
S.S. sheets. A vacuum cover was placed over the entire assembly and
evacuated prior to allowing the epoxy to set overnight. Figure~\ref{fig-lead-plate} is a
photo of a S.S. plate being placed onto a layer of lead panels
that have been painted with black epoxy.

Once the epoxy had set overnight, a U-channel beam was
placed across the top of the completed lead panel and screwed into position. The U-channel beam was then lifted with a forklift to position the lead panel and rotated to hang vertically.
The lead panel was then moved to a rollabout cart and mounted to a frame. Once the panel was positioned and held into place by the
cart frame, the U-channel beam was removed. The completed \pod{}ule, was then lifted vertically and mounted onto the lead panel.
This process was repeated seven times to complete an ECal \spodule{}.
After the seventh \pod{}ule was added, threaded stainless steel rods were screwed
through clear holes around the periphery of the \pod{}ules and secured in place with a special flat nut. 

\begin{figure}[h]
\centering
\includegraphics[width=80mm]{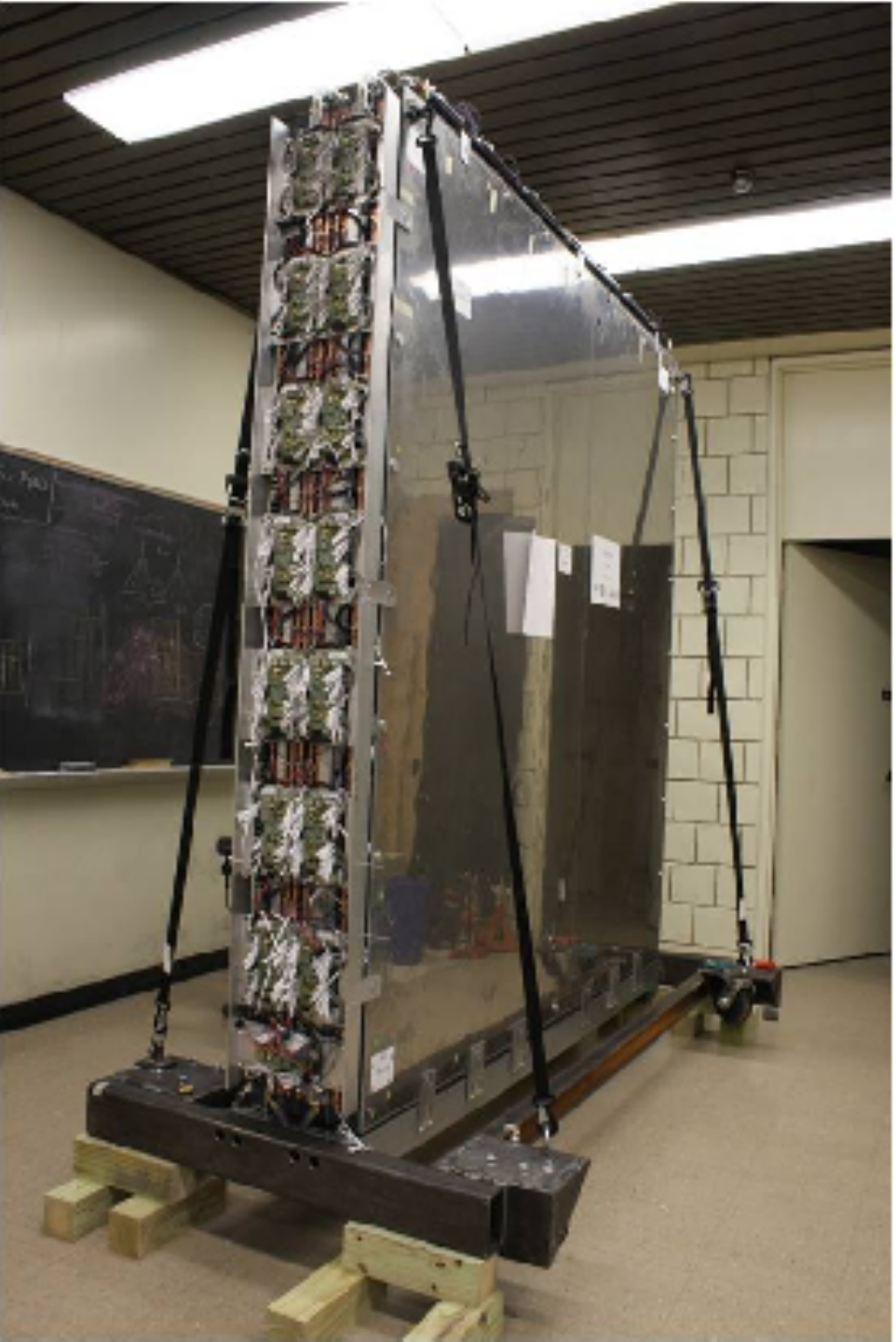}
\caption{The fully-assembled upstream ECal \spodule{} including the TFB readout electronics boards.} 
\label{fig-ecal-mounting}
\end{figure}

Finally the electronics mounting rail assemblies were attached to the top and on one side of the
ECal \spodule{}s where the MPPCs were mounted.  Once the readout electronics (see Section~\ref{sec_electronics}) were installed, the mini-coax cables from 1820 MPPCs were connected to the TFBs and labeled, completing the assembly.  Figure~\ref{fig-ecal-mounting} is a
photo of a fully assembled \pod{}ule, mounted vertically on a rollabout stand.

\subsection{The Water Target Super-P\O{}Dule Assembly}
\bigskip

The water target bag \spodule{} assembly on the \pod{}ules consisted of two water
bladders that were mounted on the face of one \pod{}ule and encased by HDPE
frames on four sides (top, bottom, left, and right). This procedure was
performed on 25 \pod{}ules to form two separate water target \spodule{}s that
held 12 or 13 water layers.

The assembly began with a \pod{}ule placed flat on an assembly table. Two
brass sheets of thickness, 0.15 cm (0.1 R.L.) were placed on the \pod{}ule.
A thin weather strip gasket was placed along the \pod{}ule
edges on two sides and the bottom. Three HDPE frames were then placed over
the weather strips to form a water tight gasket seal and then a vertical
HDPE center divider strip was placed in the middle of the \pod{}ule.  Next, the
two water bladders were placed on the brass sheet. Each water bag had a top and bottom HDPE frame.
The next step was to place the weather strip gasket in a groove in the three HDPE frames.

Finally a thin HDPE cover sheet was placed over the entire surface to form a
water-tight seal and to keep the water bladders from protuding out when the
\pod{}ule was mounted vertically and moved. The upper water bladder HDPE frame
was supported by slots in the upper HDPE\ gasket frame. The lower water
bladder HDPE frame had mounting holes that were screwed down with screws
extending through the bottom HDPE gasket frame. The bottom HDPE gasket frame
had drain holes or ports, which were connected to a drain hose to allow any
water leak to drain out in a controlled manner.

\subsection{Shipping and Installation of the Super-P\O{}Dules}

Each of the four \spodule{}s  was mounted on a solid wooden
base and enclosed in a wooden crate for shipping. The four \spodule{}s  were
flown from JFK airport in New York to the Narita International Airport near Tokyo by
the Nippon Express shipping company in April  2009. The crated \spodule{}s 
were then delivered to J-PARC on two flatbed trucks and offloaded
into the LINAC building where they were unpacked, removed from their
wooden bases and checked out in preparation for installation. All
electrical and plumbing utilities were installed in the detector hall
and attached to the off-axis detector basket while the \spodule{}s  were being
checked out in the LINAC building, which simplified the installation of
the \spodule{}s  into the basket.

\begin{figure}[htbp]
\centering
\includegraphics[width=0.4\textwidth]{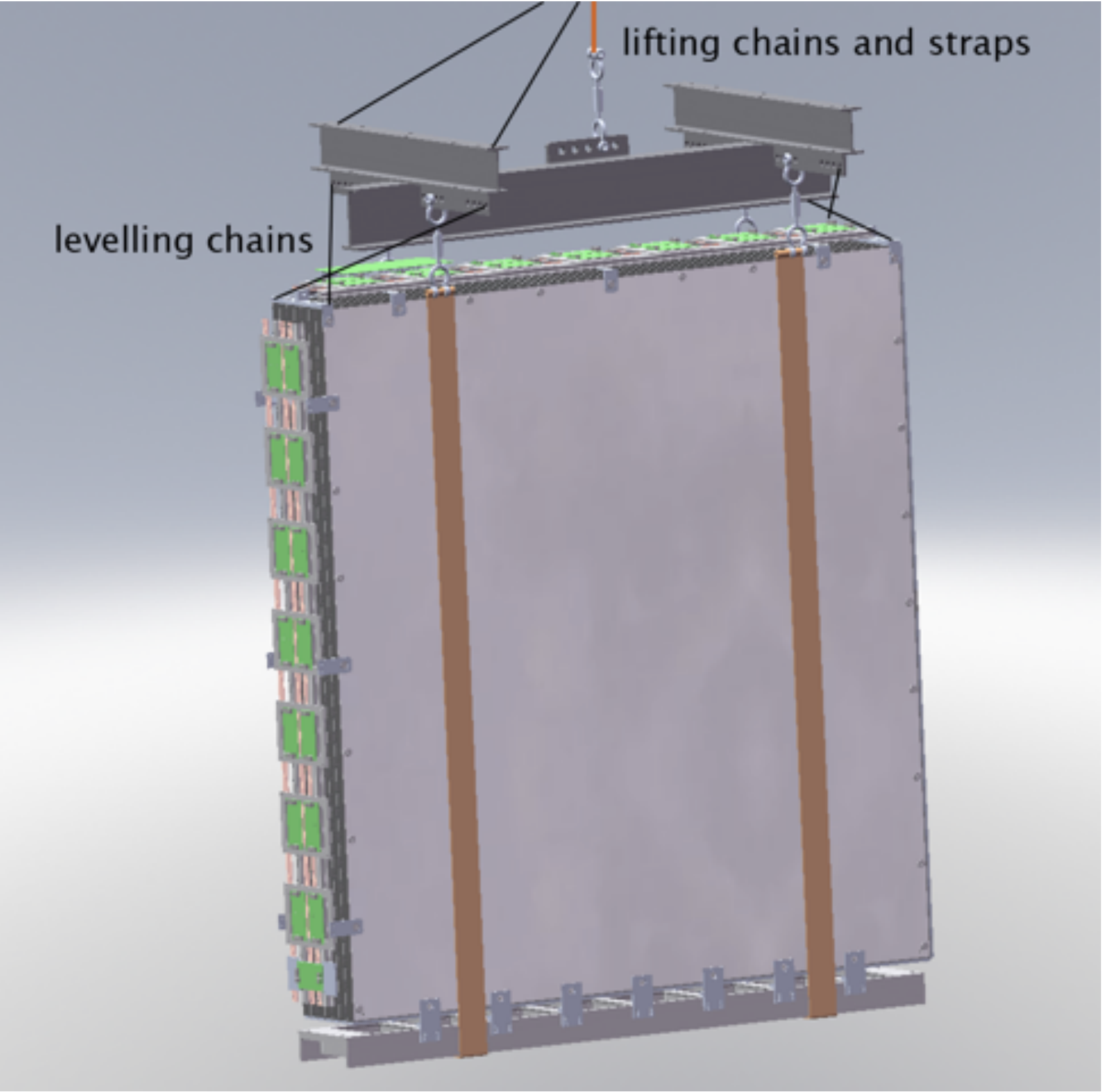}
\includegraphics[width=0.45\textwidth]{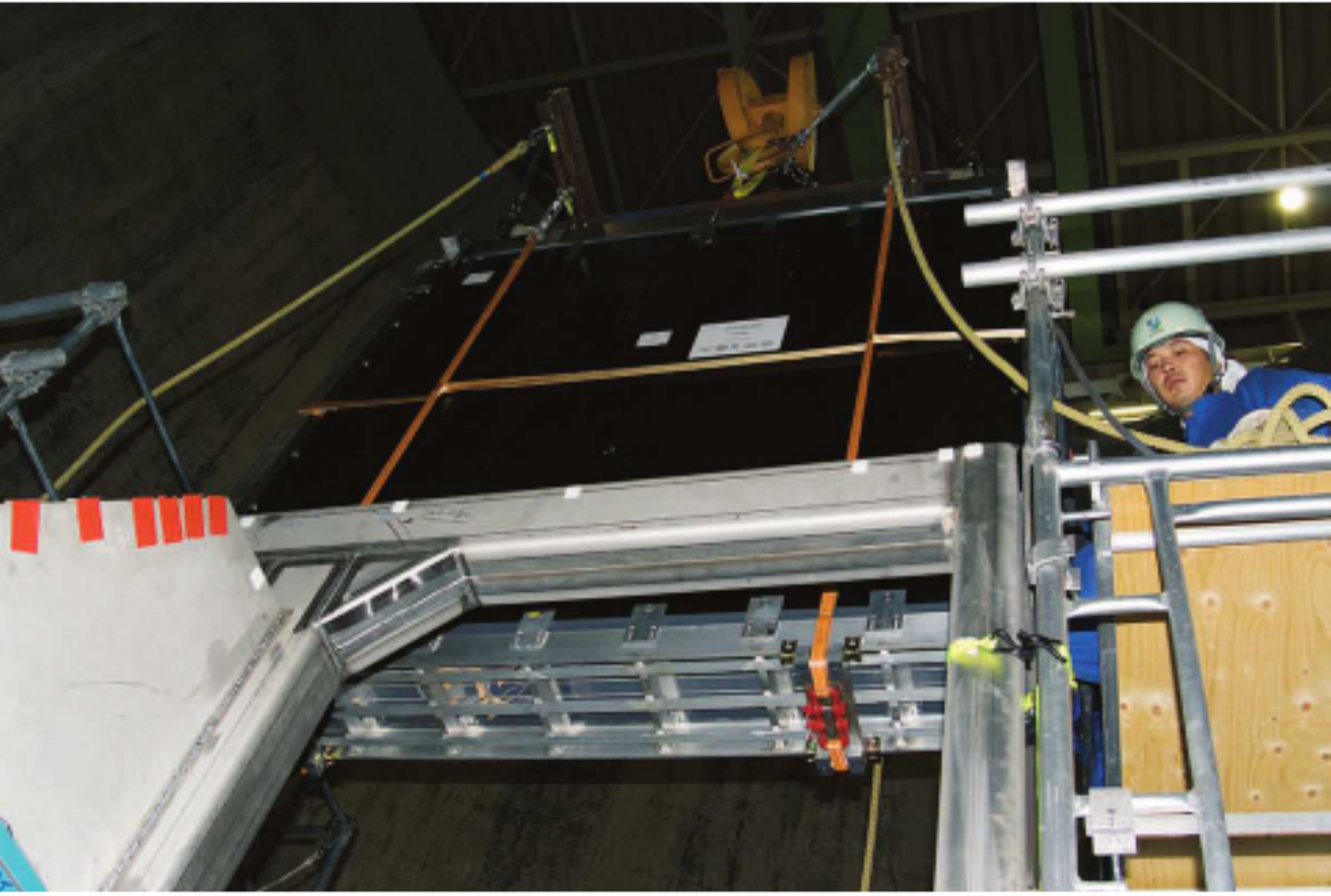}
\caption{The left figure shows an engineering drawing of an ECAL module being lifted by the custom
lift fixture, while the right picture shows the first \spodule{} (upstream ECAL) being lowered into the  detector basket.}
\label{fig:ecallift}
\end{figure}

Each \spodule{} remained bolted to custom rolling carts until just prior to
installation. Aluminum covers were attached to the side and top
electronics of the \spodule{}  to prevent damage during installation into
the detector basket. \spodule{}s  were installed in order from the most
upstream side of the detector basket. For installation each \spodule{}  was
disconnected from its cart and lifted by a crane using a custom lift
fixture (Fig.~\ref{fig:ecallift}) that allowed precise positioning
of the \spodule{}  as it was lowered into the basket about 20~m below the
staging area. Figure~\ref{fig:ecallift} shows the first \spodule{}, the
upstream ECAL, being lowered into position with guidance from a local
contractor.

Once a \spodule{}  was set into position in the basket, it was bolted into
place and utility connections were made to it. Precise positioning of
each \spodule{}  in the basket was accomplished using adjustment
screws. Following the utility connections and the filling of the water
targets, aluminum covers were attached to the \pod{} in preparation for
data taking.

\section{Readout Electronics and DAQ}
\label{sec_electronics}



\label{sec_tfb}

The \pod{}, like the SMRD, ECal, and INGRID detectors in the ND280, uses Trip-t-based front-end electronics~\cite{tfb} to read out its 10,400 photosensors. 
Each Trip-t Front-end Board (TFB) contains four 32-channel Trip-t ASIC chips~\cite{tript}, originally developed at Fermilab for the D\O{}  experiment, and can serve up to 64 MPPC sensors, which are connected by miniature coaxial cables to the TFB. 
The signal from the photosensor is divided capacitively in the ratio of 10:1 and routed into separate Trip-t channels to provide a high and a low gain response to the same input, thereby increasing the dynamic range of the electronics. 
The high gain channel provides measurement for up to a 50 photoelectron (p.e.) signal with  $\sim$10 ADC/p.e. resolution, while the low gain channel can be used to measure larger signals up to about 500 p.e.

The Trip-t chips integrate the charge in 23 consecutive integration cycles that are synchronized with the accelerator beam spill such  that each 58 ns wide beam bunch, separated by $\sim$ 580 ns, falls into a separate integration window 100 ns away from the start of each integration cycle. 
The length of the integration cycle and the reset period between them can be programmed and are set to 480 ns and 100 ns respectively for beam operation. These values will virtually eliminate deadtime for in beam spill interactions but will result in some deadtime for out of spill events such as Michel electrons.
The Trip-t stores the results of the 23 integration cycles in its analog pipeline, which are digitized using a 10-bit ADC when the data are transferred off the board following the end of the integration period.   The high gain channels also feed an internal discriminator that measures the time when the integrated signal exceeds a programmable preset threshold of about 2.5 p.e. in each integration cycle. 
The operation of the TFB is controlled by a Field Programmable Gate Array (FPGA), which also provides time stamping for the discriminator output with a 2.5 ns resolution, and moves the data from the Trip-t to the back-end electronics.

The TFB operation requires four low voltage levels.
The bias voltages  for the MPPCs are also provided through the TFB with the central core of the mini-coax cables connecting the sensors to the board being at the same uniform voltage level ($\sim$70 V). 
Individual voltage adjustments are achieved by setting the voltage level of the shield sheath of the coax cable between 0 to 5 V in 20 mV increments using an 8-bit DAC.

The entire \pod{} readout requires 174 TFBs; 29 for each of the upstream and central ECal \spodule{}s and 58 for each of the water target \spodule{}s. 
The back-end electronics of the \pod{} consist of 6 read-out merger modules (RMM), a cosmic trigger module (CTM) shared with the SMRD and the downstream ECal (DsECal), a slave clock module (SCM), and one master clock module (MCM) for the whole ND280 detector. 
Each RMM serves as a communication interface between the 29 TFBs and the data acquisition (DAQ) system, by passing control commands, clock, and trigger signal in one direction, and data in the other. 
Signals between the RMM and TFBs are transmitted using the LVDS protocol via shielded Ethernet cables, while the RMMs are connected to the DAQ computers with optical Gigabit Ethernet links. 

Cosmic trigger primitives are formed from the 29 TFBs on the upstream ECal \spodule{} based on coincidences between some number of MPPCs. 
These trigger primitives are transmitted to the CTM and combined with other trigger primitives from the SMRD and the DsECal to create a global cosmic trigger decision when any side pairs of the ND280 detector are traversed by a cosmic ray muon.
The MCM receives the accelerator timing signals when a spill happens and transmits trigger as well as periodic clock synchronization signals to the RMMs and TFBs via the SCM. 
The MCM can also generate pedestal and calibration triggers, such as for synchronous operation of the \pod{} light injection system. 
The SCM duplicates most of the functionality of the MCM and allows the configuration and operation of each sub-detector, such as the \pod{}, in standalone mode independent of  the other detectors.


The global ND280 DAQ~\cite{T2KNIM}~\cite{nd280daq} uses the MIDAS framework~\cite{midas}, developed at TRIUMF, operating on computing nodes running Scientific LINUX operating system. 
The TFBs are controlled and read out by the front-end processing nodes (FPN), each of which serves two RMMs (3 FPN are used for the \pod{}).   The read-out and configuration is provided by the read-out task (RXT), while the raw data from the TFBs are decoded and formatted for output by the data processing task (DPT). 
The DPT also performs per-channel histogramming of specific trigger types before zero suppression and inserts the histograms to the output data stream periodically. 
An event builder process collects the fragments from the sub-detectors, and writes the fully-assembled events to a buffer after basic consistency checks. 
Finally, an archiver process transfers the completed files to mass storage and creates a preview copy on a local semi-offline system for fast-turnaround calibration and data quality checks. 

A global slow control (GSC) complements the global DAQ using the same MIDAS-based software framework. 
Front-end tasks running on the sub-detector computers connect to various equipment of the sub-detectors and collect monitoring data that are stored in a MySQL database at regular intervals. 
Any of the variables collected by the GSC can be displayed conveniently through a web interface. 
Several customized web pages have been developed for controlling and monitoring different components, such as the power supply voltages, TFB internal and external temperatures, the \pod{} water target levels, etc. 
Alarms can be set interactively to catch variables out of range and to alert shift personnel.


About 210 TFBs were tested prior to installation in the \pod{}. 
The test utilized the internal calibration circuit of the boards, which injects a specific charge into the TFB input channels. 
The injected signal is provided by a 10~pF capacitor charged to a voltage level up to 5 V, specified by a 12-bit DAC, and then discharged into the selected input channels. 
The basic functionality of the TFBs was tested by measuring the response of all Trip-t channels at 45 different calibration levels. 
An RMM emulator board and a simplified version of the MIDAS based DAQ was used to operate the test system and collect the data. 
The boards with the most uniform gains were selected for installation in the \pod{} while the others were set aside as spares. 
The spread of the electronics gain measured in the tests was consistent with the 5\% tolerance of the charge division and calibration capacitors used in the manufacturing of the TFBs.

\subsection{Electronics mounting and cooling}

The power requirements for the 174 TFBs and 6 RMMs in the \pod{} are listed in 
Tables~\ref{table:TFBpwr},~\ref{table:Spodpwr} and~\ref{table:RMMpwr}.
The TFBs were mounted on six rails
attached to the outside face of the \pod{}, as shown in Fig.~\ref{fig:3dp0d}. 
Each rail consists of two aluminum
extruded pipes that contain the cooling water at negative pressure. Aluminum plates, 
6.4 mm $\times$ 300 mm $\times$ 203 mm, were attached directly to 
the extrusions. The TFBs are thermally mounted to the plates using 
Dow Corning TP-2160-T4.6-5002 thermal pads. 
Two TFBs are mounted per plate and there are seven plates of this type in 
each view. The $y$ coordinate readout has an extra plate with only one TFB mounted. 
Therefore one rail has a total of 29 mounted TFBs. One RMM services each rail and 
is mounted on additional aluminum plates attached to the TFB rails.
The upstream and central ECals each employ a single rail while the upstream and central
water targets each require two.

The voltage requirements of the TFBs including the MPPC voltage are 
5.5 V, 3.8 V, 3.1 V, 1.7 V, and 70 V(HV). Power and ground are 
supplied to the TFBs via six copper bus bars (3.2 mm $\times$ 12.7 mm for power 
and 6.4 mm $\times$ 12.7 mm for ground) traversing the full length of the rails. 
The voltages and ground are supplied by 22 AWG wire coming from the TFB power connector (MOLEX 90142-0020) jumpers to these bars. 

\begin{table}[ht]
\begin{center}
\begin{tabular}{|c|c|}
\hline
\multicolumn{2}{|c|}{Power Requirements per TFB} \\
\hline
\hline
Voltage (V) & Current (A) \\
\hline
1.7 & 0.26 \\
3.1 & 0.60 \\
3.8 & 0.05 \\
5.5 & 0.18 \\
\hline
\end{tabular}
\end{center}
\caption{Power requirements per TFB. A total of 174 TFBs were needed.}
\label{table:TFBpwr}
\end{table}

\begin{table}[ht]
\begin{center}
\begin{tabular}{|l||c|c||c|c|c|}
\hline
\multicolumn{6}{|c|}{Power Requirements per \spodule{}} \\
\hline
\hline
\spodule{} & \# & \#  & Vol. & Cur. & Power \\
& TFB & RMM &  (V) & (A) & (W) \\
\hline
Upstream&&&&&\\
ECal&29&1&1.7 & 7.5 & 12.8\\
&&&3.1 & 19.9 & 61.7\\
&&&3.8 & 4.4 & 16.8\\
&&&5.5 & 5.7 & 31.4\\
\hline
Upstream&&&&&\\
Water Target&58&2&1.7 & 15.0 & 25.6\\
&&&3.1 & 39.8 & 123.4\\
&&&3.8 & 8.8 & 33.6\\
&&&5.5 & 11.4 & 62.8\\
\hline
Central&&&&&\\
Water Target&58&2&1.7 & 15.0 & 25.6\\
&&&3.1 & 39.8 & 123.4\\
&&&3.8 & 8.8 & 33.6\\
&&&5.5 & 11.4 & 62.8\\
\hline
Central&&&&&\\
ECal&29&1&1.7 & 7.5 & 12.8\\
&&&3.1 & 19.9 &61.7 \\
&&&3.8 & 4.4 & 16.8\\
&&&5.5 & 5.7 & 31.4\\
\hline
\hline
\bf{Total}&&&&&736.2\\
\hline
\end{tabular}
\end{center}
\caption{Power requirements per \spodule{}}
\label{table:Spodpwr}
\end{table}

\begin{table}[ht]
\begin{center}
\begin{tabular}{|c|c|}
\hline
\multicolumn{2}{|c|}{Power Requirements per RMM} \\
\hline
\hline
Voltage(V) & Current(A) \\
\hline
3.1 & 0.37\\
3.8 & 2.88 \\
5.5 & 2.44 \\
\hline
\end{tabular}
\end{center}
\caption{Power requirements per RMM. A total of 6 RMM were needed.}
\label{table:RMMpwr}
\end{table}


\section{Response Scan of the P\O{}Dules}




\subsection{Construction of the P\O{}Dule Scanner}

After assembly of each \podule{}, the response of each scintillator
bar was measured using an automated radioactive source scanner.  The
goal of the source scan was to measure the position of each bar in the
\podule{}, to measure variation of response of
each bar along its length and to check for dead or otherwise
compromised channels.

\begin{figure}[htbp]
\centering
\includegraphics[width=0.47\textwidth]{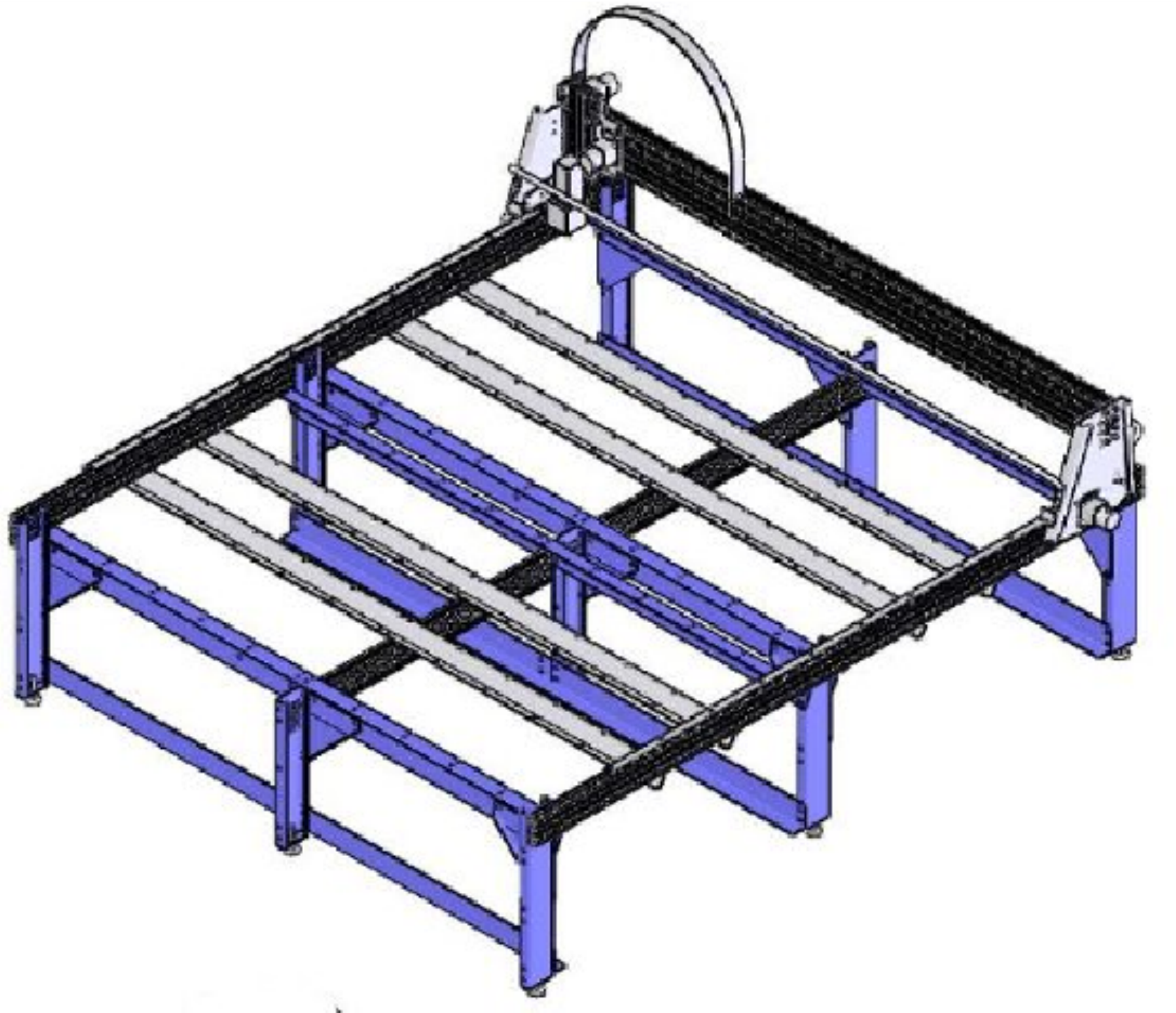}
\includegraphics[width=0.47\textwidth]{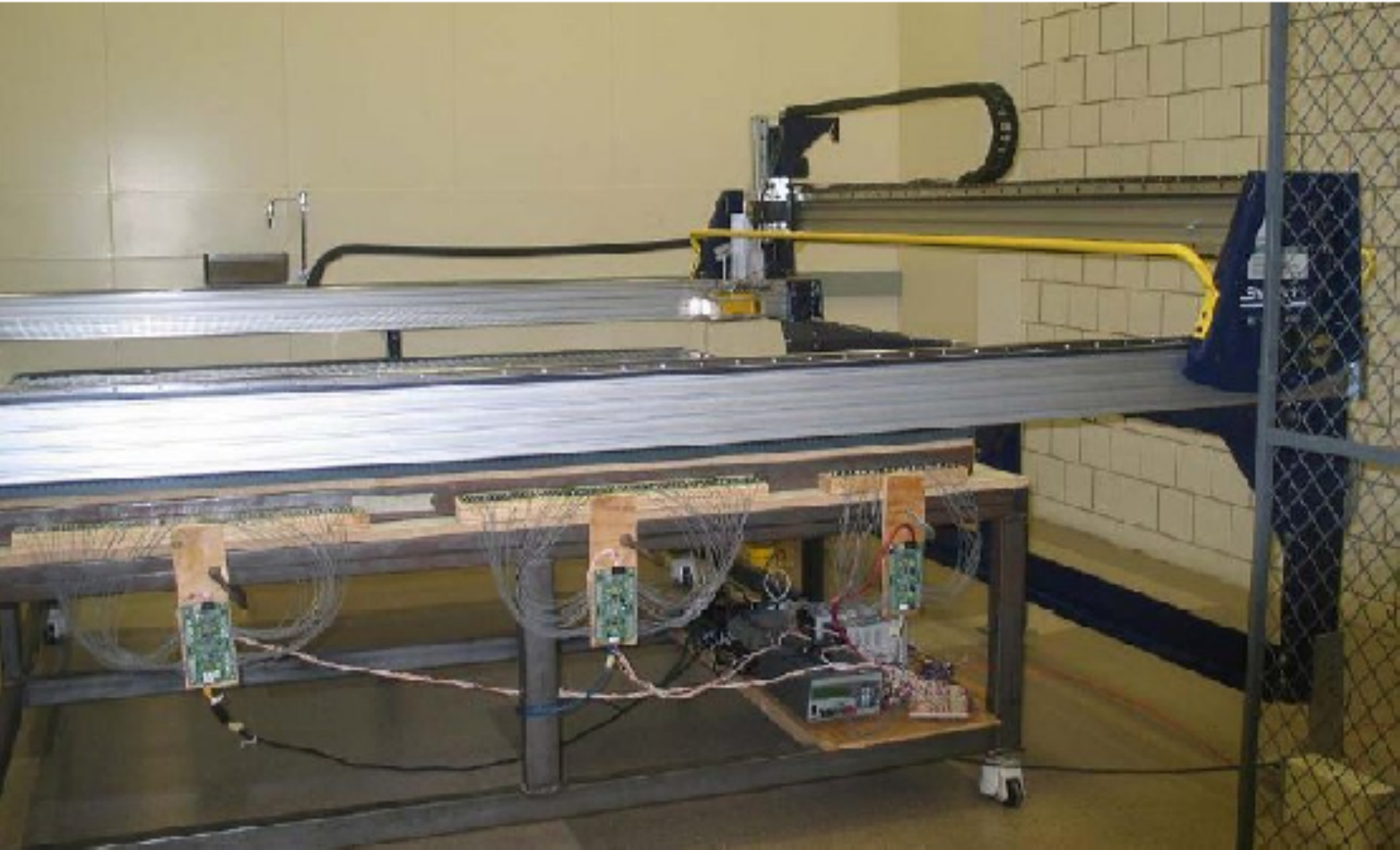}
\caption{An engineering schematic (left) of the custom CNC router modified to
  serve as the \pod{} scanner (right), shown with a \podule{} that is ready to be scanned.}
\label{fig:scanner}
\end{figure}

The scanner was based on a $9' \times 10'$ custom CNC router manufactured by
ShopBot of Durham, North Carolina.
To convert it for use as a scanner, the spindle router was replaced with a unit
consisting of a radioactive source, a video camera for alignment of the
\podule{} to the router table coordinates, and a depth sensor to
measure distance from the source to the surface of the \podule{}.  The
source was a $0.7$~mCi $^{60}$Co source housed inside a $4$'' thick
lead shield.  Inside the shield, the source was located in a $0.27$''
diameter bore hole, $1.75$'' from the bottom surface, and collimation
of the source on the \podule{} was achieved by cutting away the lead
in a fixed cone with the curved surface angled $22.7^{\circ}$ from the vertical.
Figure~\ref{fig:scanner} shows the scanner with a \podule{}
ready to be scanned.  

The scanner motion was controlled by a LabView based program to drive
the source to predetermined locations so that a current in the photosensor due to the
source could be measured on each channel.  For each of the two views,
the source was moved perpendicular to the bar direction in $1.7$~cm
steps.  Since the beam from the source was much wider than the step
size, response as a function of position could determine both the
location of the strip and its response to the source at that point.
Eleven such scans were performed at different positions along the
strips to measure the change in response along the bar.

\subsection{Operation of the Scanner}
\label{sec_scanner}

Each scan was conducted in 1.7 cm steps along eleven equally spaced lines for both $x$ and $y$ coordinates. The P\O{}Dule corners were used as reference points for the scanning and viewed using a web camera mounted on the moving scanner element next to the source shielding. The scanner control and DAQ computer provided the coordinates for the next scanning position in addition to collecting and processing the data. The same set of readout electronics (See Section~\ref{sec_electronics}) was used for all scanning operations.

\subsection{Analysis of Scanner Data}

\begin{figure}
\includegraphics[width=8cm]{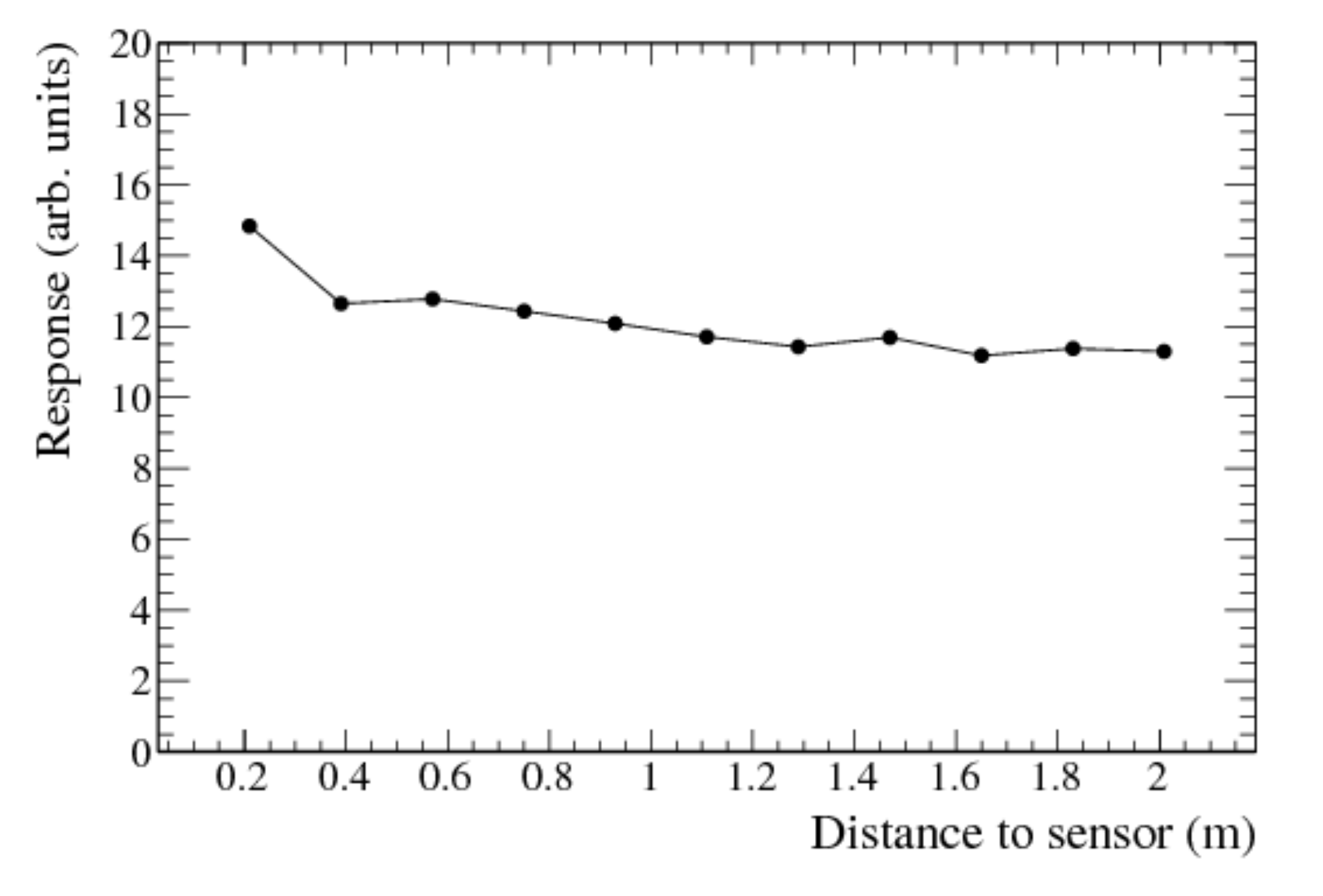}
\centering

\caption{Typical scanner response of a single scintillator bar.}
\label{fig:scanner_1bar}
\end{figure}

The manufacturer-suggested bias voltage values were used for each sensor without any additional tuning.  Since a gamma radioactive source was used for the scanning process, a random trigger was used. First, the dark noise spectrum  was obtained for the MPPC sensors without the source, which established the pedestal and provided an independent check of the MPPCs.  A channel exposed to the source experiences a pedestal shift to lower values proportional to the exposure. Thus the channels with source signal will have the pedestal at lower ADC counts than without it. It was expected that the 11 points measured along each bar should lie on an approximately exponentially decaying smooth curve corresponding to the light attenuation in a WLS fiber \cite{Beznosko:2007mm} with a mirrored end. A sample of the scanner output for a single scintillator bar is given in Figure~\ref{fig:scanner_1bar}. A sharp step in this curve would indicate damage along the optical fiber. Only two such fibers were discovered (and replaced), out of 10,400 \pod{} channels. 

Additionally, the spread of the channel outputs was obtained.  For example, for the $x$ \podule{} bars, as the scanner moved along the $x$-axis with a fixed $y$ position, the output of each channel was recorded as a function of $x$. For each bar this resulted in a constant pedestal value with a peak when the scanner reached the $x$ position of that bar.  A Gaussian was fit to that peak, to determine the response of the bar at that $y$ position. This was then repeated for a number of different $y$ positions along the bar.

\begin{figure}
\includegraphics[width=8cm]{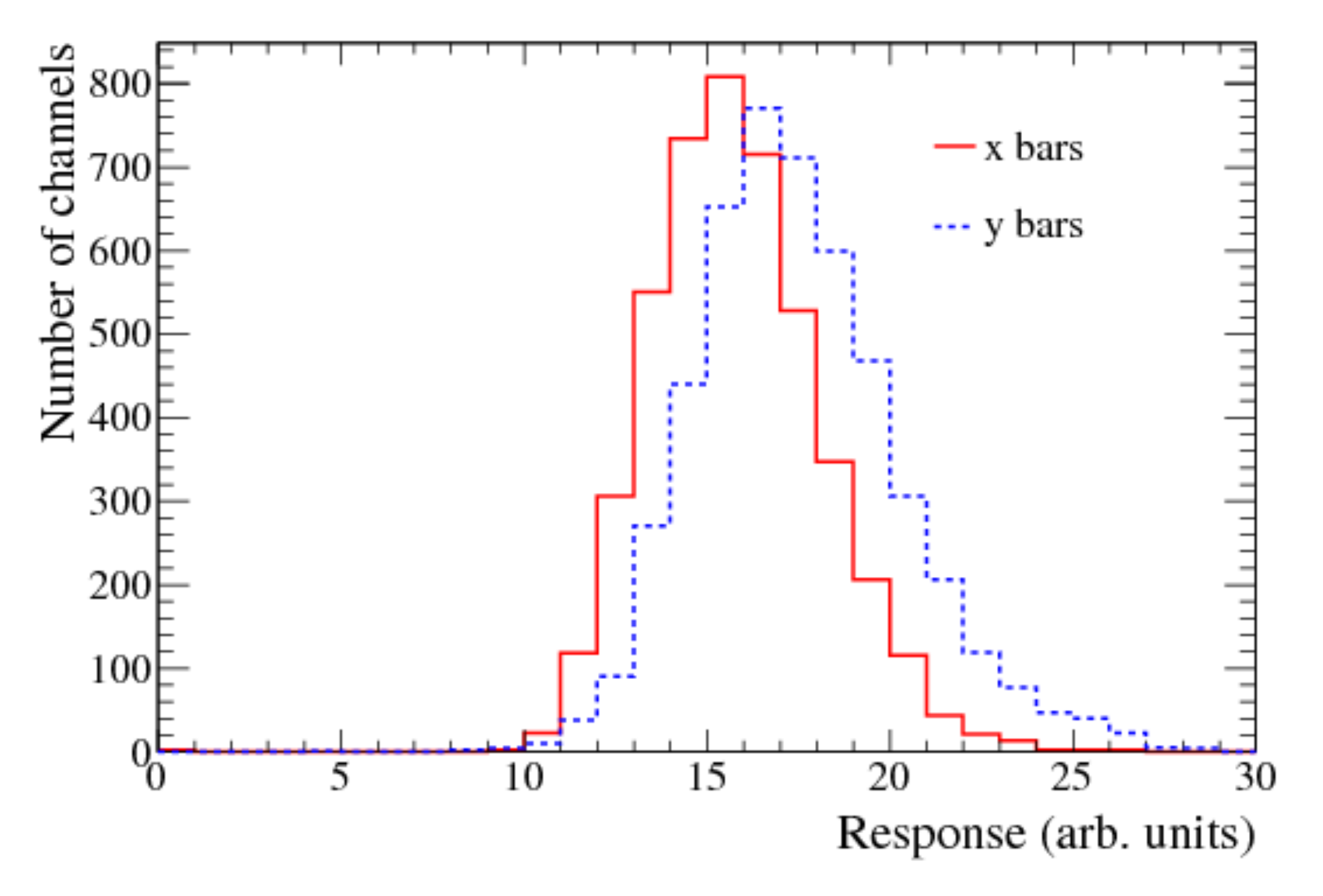}
\includegraphics[width=8cm]{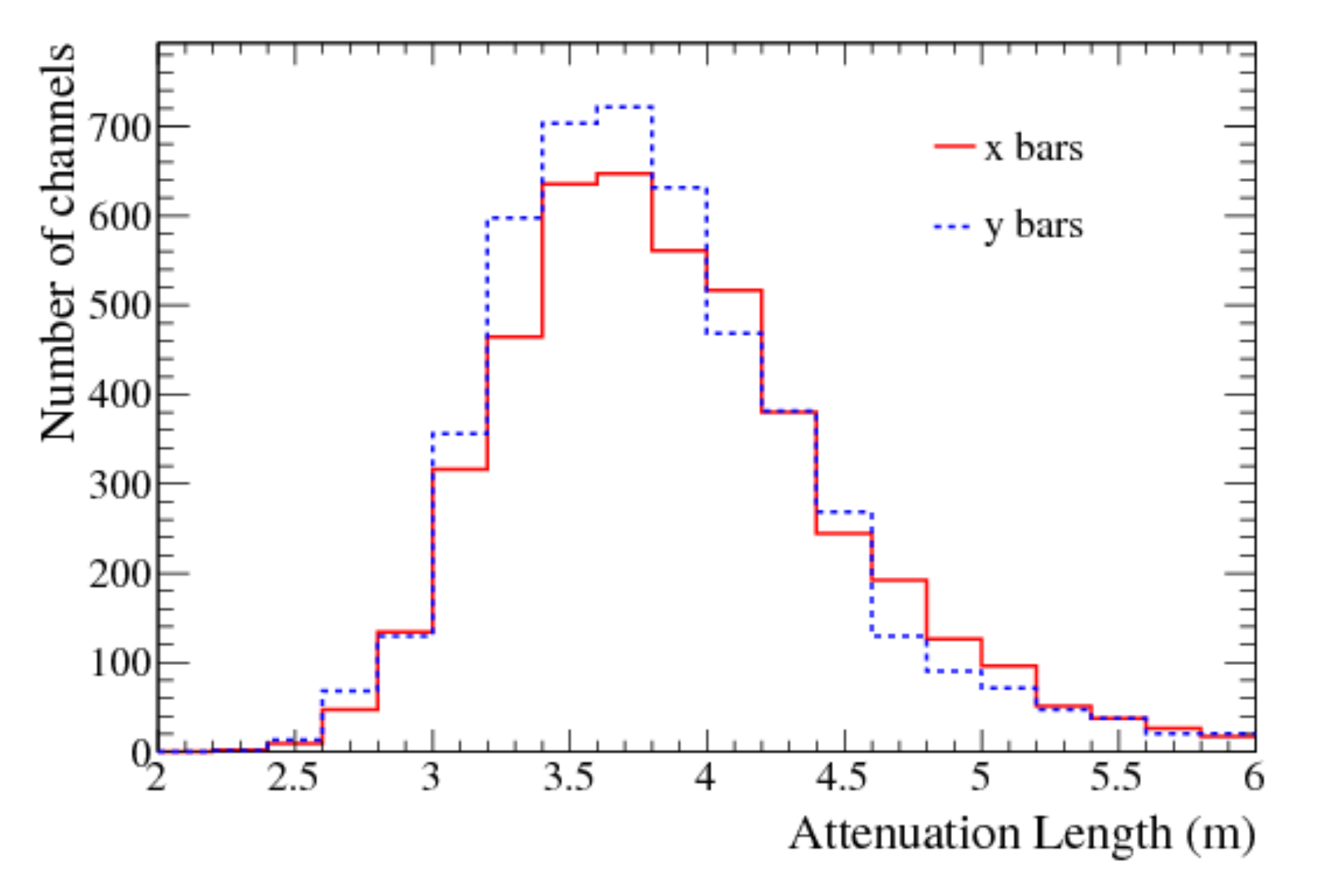}
\centering
\caption{ Corrected response ($N$ in Equation~\ref{eq:scanner_atten}) for both $x$ and $y$ bars (top); Long attenuation length ($L$ in Equation~\ref{eq:scanner_atten}) for both $x$ and $y$ bars (bottom).}
\label{scanneroutput_figure1}
\end{figure}

The bar response was fit as a function of position along the bar to a double exponential attenuation, motivated by the known two dominant WLS emission peaks, plus a reflection term, namely

\begin{equation}
N\times [ fe^{\frac{-y}{L}} + (1-f)e^{\frac{-y}{S}} + R(fe^{\frac{-(2b-y)}{L}} + (1-f)e^{\frac{-(2b-y)}{S}} )] ,
\label{eq:scanner_atten}
\end{equation}
where $N$ is the overall normalization, $f$ is the fraction of light in the long mode, $L$ is the long attenuation length, $S$ is the short attenuation length, $R$ is the reflectivity, and $b$ is the length of the bar.  Here only $L$ and $N$ were allowed to vary in the fit, while the other parameters were held at fixed values.  A similar procedure was also applied to $y$ \podule{} bars.  The corrected response  and the long attenuation length  of each bar are plotted in Figure~\ref{scanneroutput_figure1}.    Note that the $y$ bars were on the top of the scanned \podule{}s, so the average response was lower in the $x$ bars due to the absorption of the source.  
 


\section{Pre-installation Testing}




\subsection{Channel Checks using Dark Noise}
\label{sec_testing_dark}

The dark noise of the MPPC detectors makes a good parameter for a quick check of the photosensor and the readout electronics channel.  The test setup consisted of three back-end boards: a local Master  Clock Module (MCM),  which provided a
synchronized  clock to the  other back-end  boards and  controlled the
data  taking;  an  RMM connected to all of the  mounted TFBs;  and a cosmic trigger module  (CTM).  For the ECal \spodule{}s, each
TFB  was  also  connected to  the  CTM,  separated  into four  groups  of
primitives (upstream $x$, downstream  $x$, upstream $y$ and downstream
$y$). 

The pre-installation check of all \pod{} channels into the ND280 detector was carried out in two steps.  The first step included dark noise testing that took place during the scanning.  The results of the dark noise test were thus indicative of the actual MPPC performance after their installation into the detector modules. 14 MPPCs were found to be outside our selection criteria and had to be replaced at this stage. The typical bad sensor signature was either no output (no pedestal or no clearly separated p.e.\ peaks on the response histogram), or an output with barely separable p.e.\ peaks. The testing was done at the manufacturer recommended biasing voltage values for each sensor, which results in a gain of around $6.5\times10^5$ at a photosensor temperature of 25$^{\circ}$ C.
     
The second step for this pre-installation check was performed after all the TFB boards were installed on the sides of each of the four Super-P\O{}Dules and had been shipped to the JPARC facility. A separate readout setup, including a portable RMM module, power supplies, and DAQ computer was connected to each TFB board. The dark noise histograms from a random trigger were plotted for each channel.
     
Dark noise checks resulted in three TFBs and two MPPCs being replaced due to anomalous dark noise rates. 
A second-pass of dark noise testing was undertaken after installation of the water system level sensors. This was deemed necessary since some electronics had been moved to allow access to the water target region. The check revealed that several coaxial cables between the MPPCs and the TFB boards had to be reconnected or replaced. After this was completed, all 10,400 P\O{}D channels were operational.

\subsection{Checkout with the Light Injection System}

Prior to installation of the \pod{} into the ND280 basket, the light injection system was used to confirm that the fiber to 
photosensor to TFB to RMM detection chain was correct functioning.

The Light Injection System (LIS) was flashed at two amplitudes: a lower amplitude to test the system response in a region 
where most of the channels were read out from the high gain ADC and a higher amplitude to test the response of the low gain ADC.  

The LIS allowed the identification of production defects that could be repaired before \pod{} installation. One method utilized was to look at the measured signal 
on all fibers within a single \pod{}ule illuminated by a single LIS pulser channel. The overall features of this distribution are understood by the 
LIS cavity geometry and the position of the fibers with respect to the light source. Anomalous signals on a fiber may indicate a production defect
such as a misplaced fiber guide.


\subsection{Checkout using Cosmic Rays}

During the \pod{} checkout prior to installation in the experimental hall,  cosmic ray data were taken with the full
data  acquisition chain and the full  software chain.  A detailed description of this work can be found in~\cite{TrungThesis}.

Each \spodule{}  was tested  independently, with the  mounted TFBs
connected to the setup described in~\ref{sec_testing_dark}.  

In the cosmic ray muon mode,  a TFB with integrated charge over threshold produces a signal sent to the CTM, which triggers the RMM to read out all of the TFBs if at least one board in each primitive group (upstream vertical bars, downstream vertical bars, upstream horizontal bars and downstream horizontal bars) produced a signal to ensure that the cosmic had passed through the \spodule{} and produced hits in both views.

The cosmic  data was  calibrated using the  dark noise data  and
passed  to  the  \pod{}  reconstruction  software.
Integration cycles  with more than  20 hits (a normally  incident muon
produces $\sim$28 hits), were  searched for muon tracks. Events with
a single track were selected and used for the first
minimum-ionizing particle (MIP) light yield study (Section~\ref{sec:MIPLightYield}).

\begin{figure}[hbt]
\centering
\includegraphics[width=8cm]{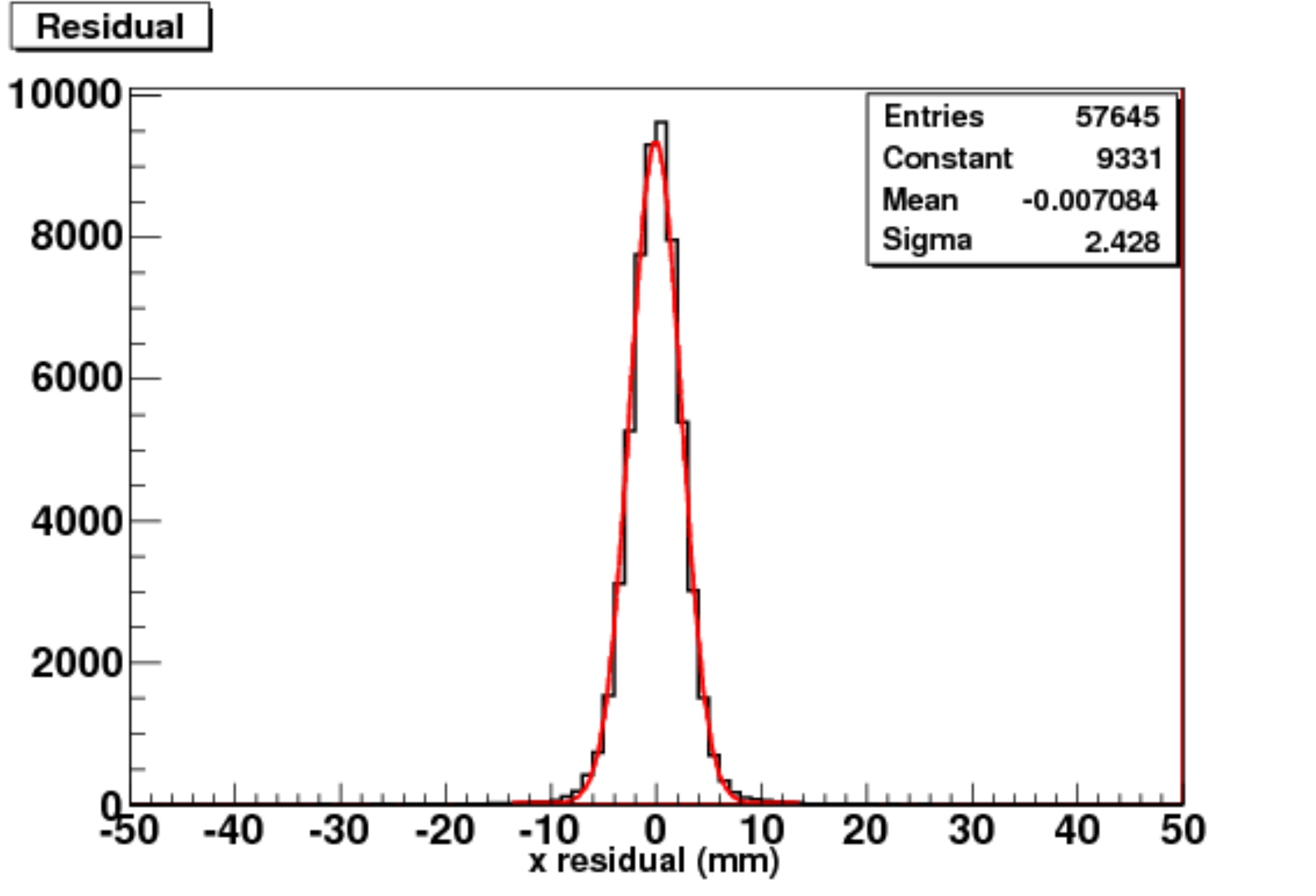}
\hspace{0.5cm} 
\centering
\includegraphics[width=8cm]{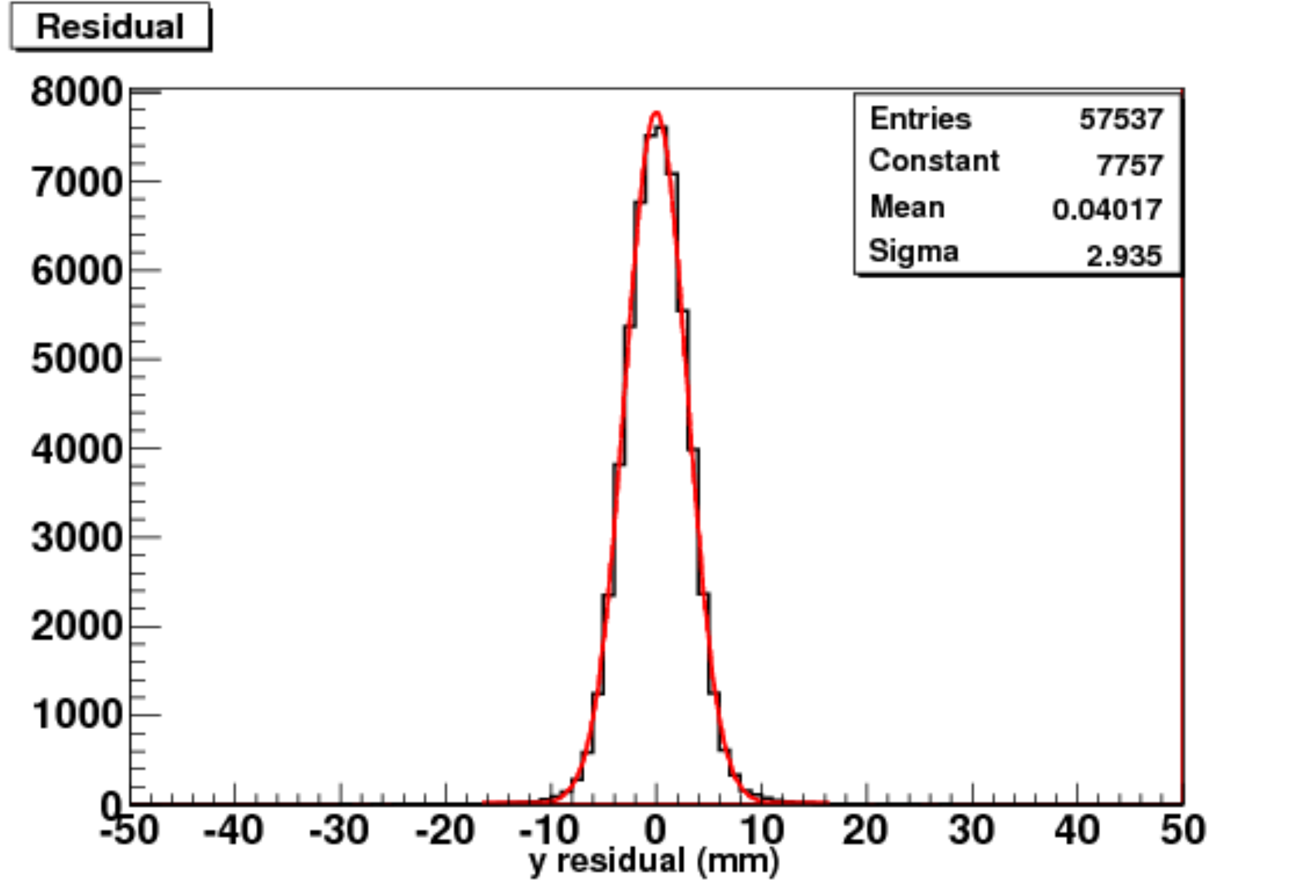}
\caption{Residual distributions of reconstructed 3D tracks (solid histogram) in the $xz$ (left) and $yz$ (right) planes. The distributions are fit to a Gaussian distribution (solid curve).}
\label{fig:residuals}
\end{figure}

Figure~\ref{fig:residuals} shows the residuals of the reconstructed
tracks  in the Upstream  ECal, which  have been  fit to  a Gaussian
distribution.   The  fitted   parameters  of $\mu_{x,y}  =  0.0$~mm, 
$\sigma_{x} = 2.8$~mm and $\sigma_{y}  = 3.2$~mm, gave the first
direct measurement of the \pod{}'s tracking resolution.

\section{Detector Calibration and Performance}




\subsection{Gain Calibration}

The low level charge calibration converts raw ADC response of the electronics to photoelectron units. It is performed in three stpdf: pedestal subtraction, correction for the electronics non-linearity and the relative low/high gain response, and correction for the MPPC gain variations.

\begin{figure}[htb]
\centering
\includegraphics[width=0.5\textwidth]{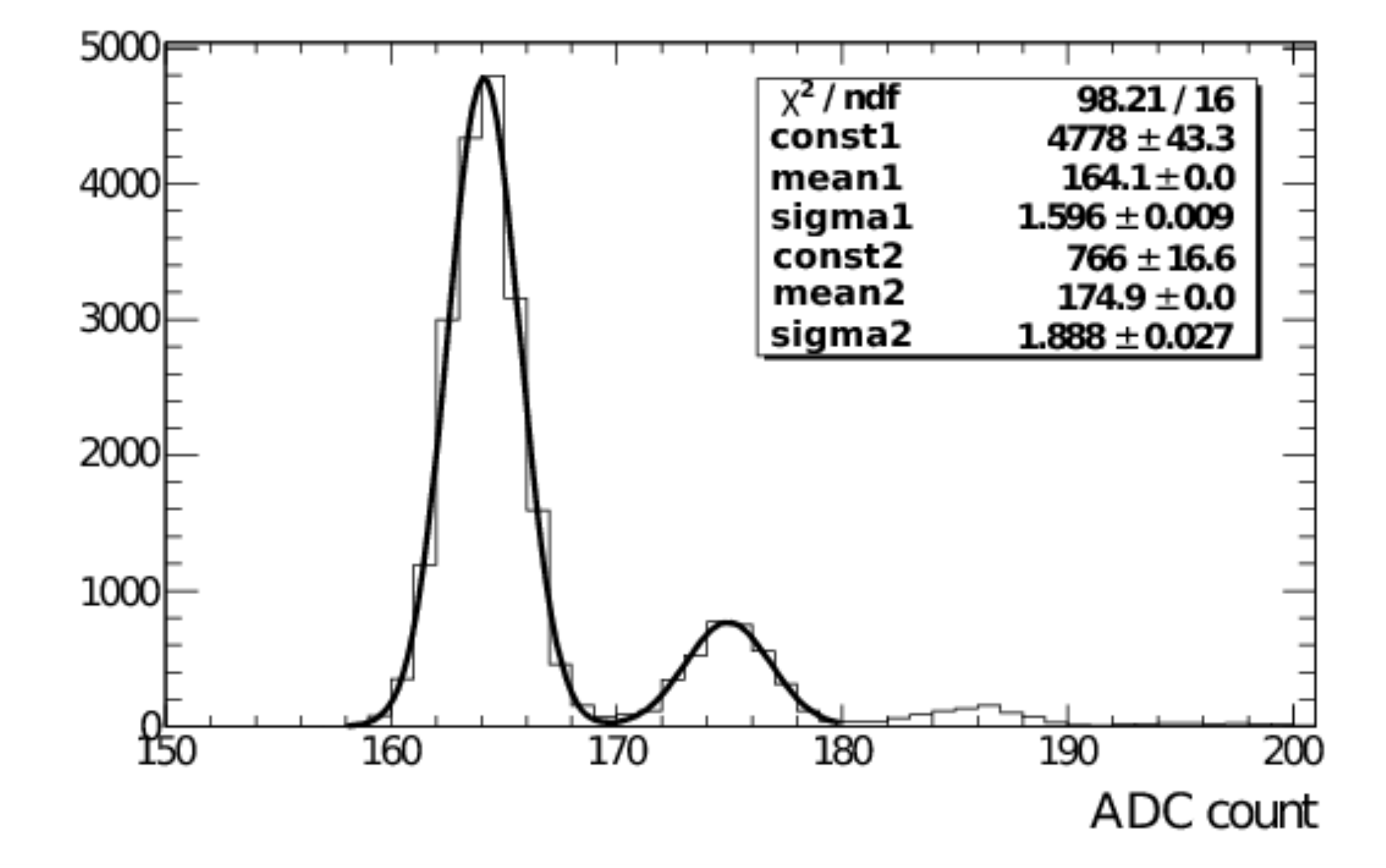}
\caption{Typical digitized dark noise spectrum of an MPPC with a double Gaussian function fitted to the pedestal and 1 p.e. peaks.}
\label{fig:dark-noise}
\end{figure}

The pedestal, i.e.~the baseline response of the MPPC and electronics without input signal, as well as the MPPC gain, is measured using non zero-suppressed dark rate noise (Figure~\ref{fig:dark-noise}). The pedestal peak  in the dark noise spectrum is fit to a Gaussian function in each integration cycle separately to account for the small variations among the cycles. The mean of the Gaussian gives the pedestal constant used for the pedestal subtraction. The MPPC gain is measured as the separation between the pedestal and the 1 p.e.\ peak after combining the dark noise spectra from all 23 integration cycles which were corrected for individual pedestal shifts. The two peaks are fit to a double Gaussian and the difference in their means is used to measure the photoelectron unit in ADC values. 
Since the MPPC overvoltage, as well as the pedestal, is quite sensitive to the temperature at a fixed bias voltage, the gain and pedestal require continuous monitoring and updating.

\begin{figure}[htb]
\centering
\includegraphics[width=0.5\textwidth]{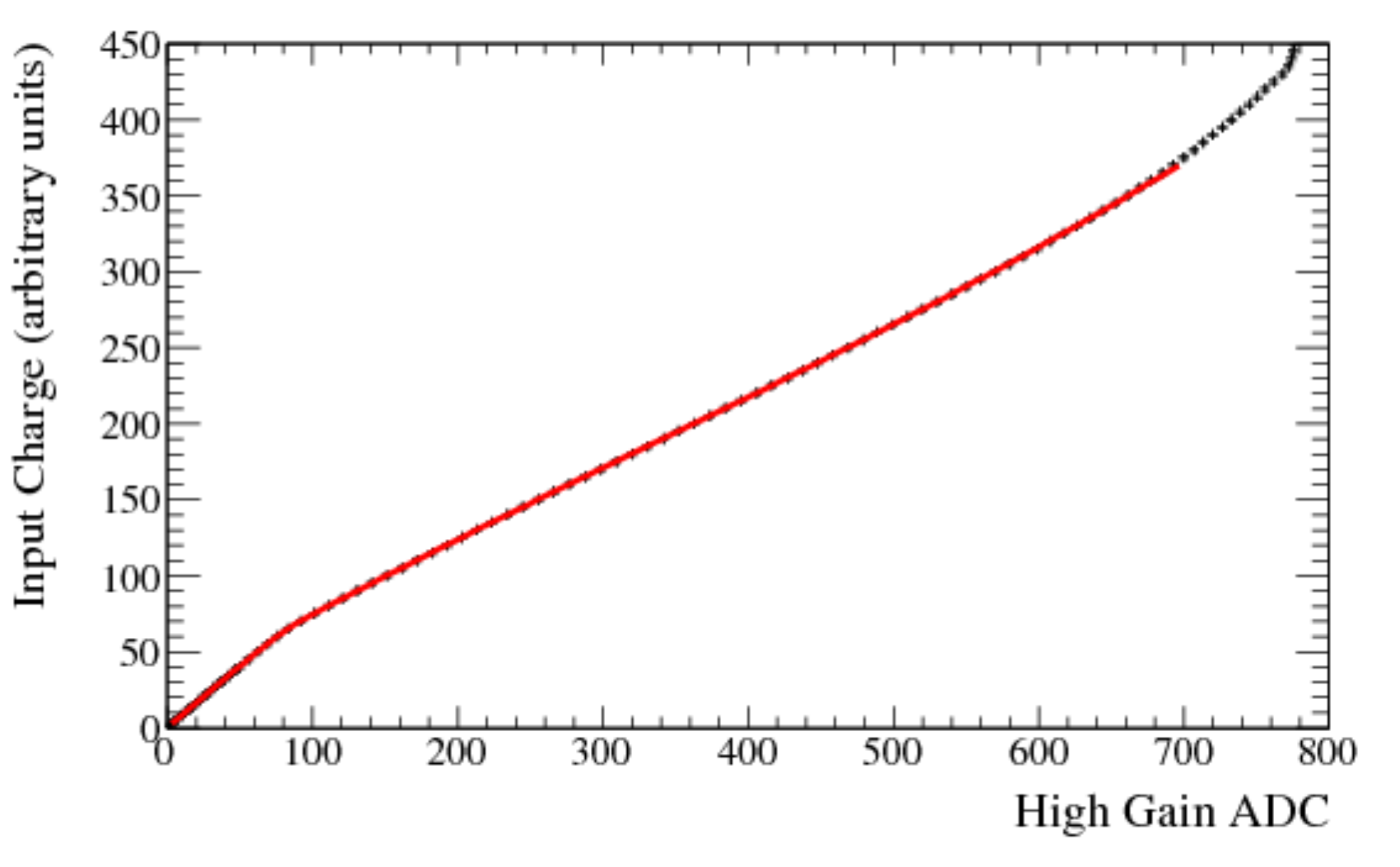}
\caption{Charge versus ADC for a high-gain Trip-t channel fit to a bi-cubic function.}
\label{fig:tript-nonlin}
\end{figure}

Before converting the signal into photoelectron units, the raw ADC response needs to be corrected for the non-linearity of the electronics and the relative gain difference between the high and low gain response. The response of each input channel is measured using the internal TFB calibration circuit as a series of 174 signal levels that covers both the high and low gain dynamic range. This measurement is performed \emph{in situ} when the MPPCs are powered since the capacitance of the photosensors and the mini-coax cables connecting the sensors to TFBs represent a significant additional capacitance on the input, altering the effective electronics gain. The measured response as a function of the calibration level is fit to a bi-cubic polynomial with nine free parameters. This parametrization is used to correct the raw ADC values during the offline calibration of the data. The bi-cubic function allows an adequate representation of the Trip-t non-linearity with residuals typically smaller than a few percent (Figure~\ref{fig:tript-nonlin}). The electronics gain and non-linearity are fairly stable, requiring only occasional checks, and therefore the constants are updated only if there is a hardware change to the front-end electronics.

\subsection{MIP Light Yield}
\label{sec:MIPLightYield}
A useful  characteristic of minimum ionizing particles  (MIPs) is that
their  energy   loss  is  only  weakly  dependent   on  their  energy.
Therefore,  for high energy  muons passing  through the  detector, the
mean energy deposition  per unit length is a  constant.  With a sample
of through-going  muons, this constant  can be determined.   The first
sample  used  was the  cosmic  checkout  data,  selecting tracks  with
$\cos\theta  > 0.8$~\cite{TrungThesis} (where $\theta$ is the angle that the track makes with the $z$ axis of the detector)  but this  was  for individual
Super\pod{}ules only.

After the  \pod{} was installed in  the basket, it  became possible to
calibrate all  \pod{}ules with the  same data sample.  The best  sample was
through-going  muons  from  beam  neutrino  interactions  in  the  wall or sand and rock
upstream of  the \pod{}.   After reconstruction, events  were selected
with a single  3D track entering the front face  of the \pod{} and
exiting out the downstream end.

\begin{figure}[htb]
\centering
\includegraphics[width=0.5\linewidth]{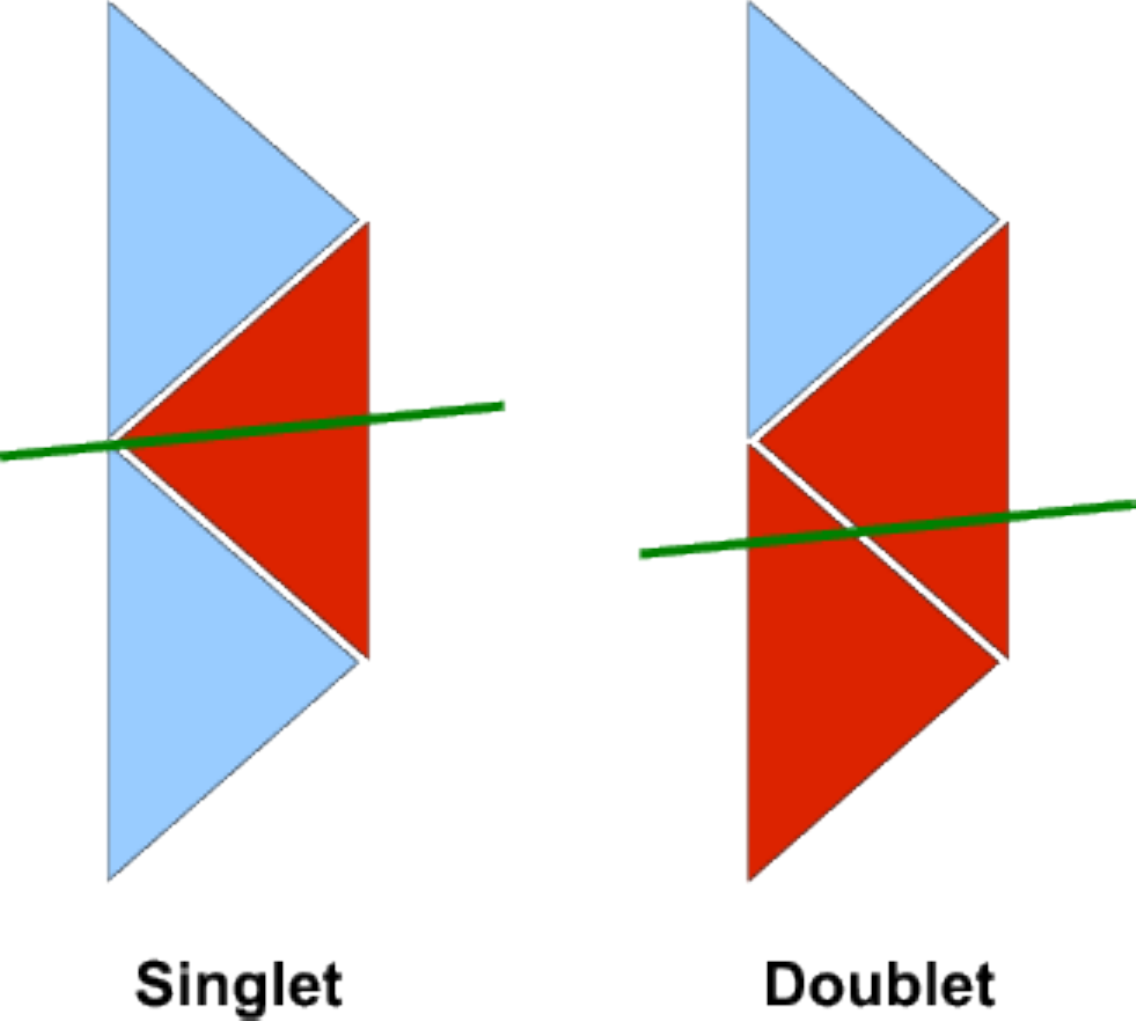}
\caption{Illustration  of  a singlet  and  doublet,  as  a MIP  passes
  through a \pod{}ule layer.}
\label{fig:mip-bars}
\end{figure}

These events were analyzed  to show each layer's detection efficiency.
Due  to the  triangular design  of  the \pod{}'s  scintillator bars,  a
normally  incident MIP is  most likely  to pass  through two  bars, as
demonstrated in Fig.~\ref{fig:mip-bars}.   However, depending on the
path taken, there  is a chance that one bar is  untouched, or that the
signal is below the noise threshold cut applied by the reconstruction.

\begin{figure}[htb]
\centering
\includegraphics[width=1.0\linewidth]{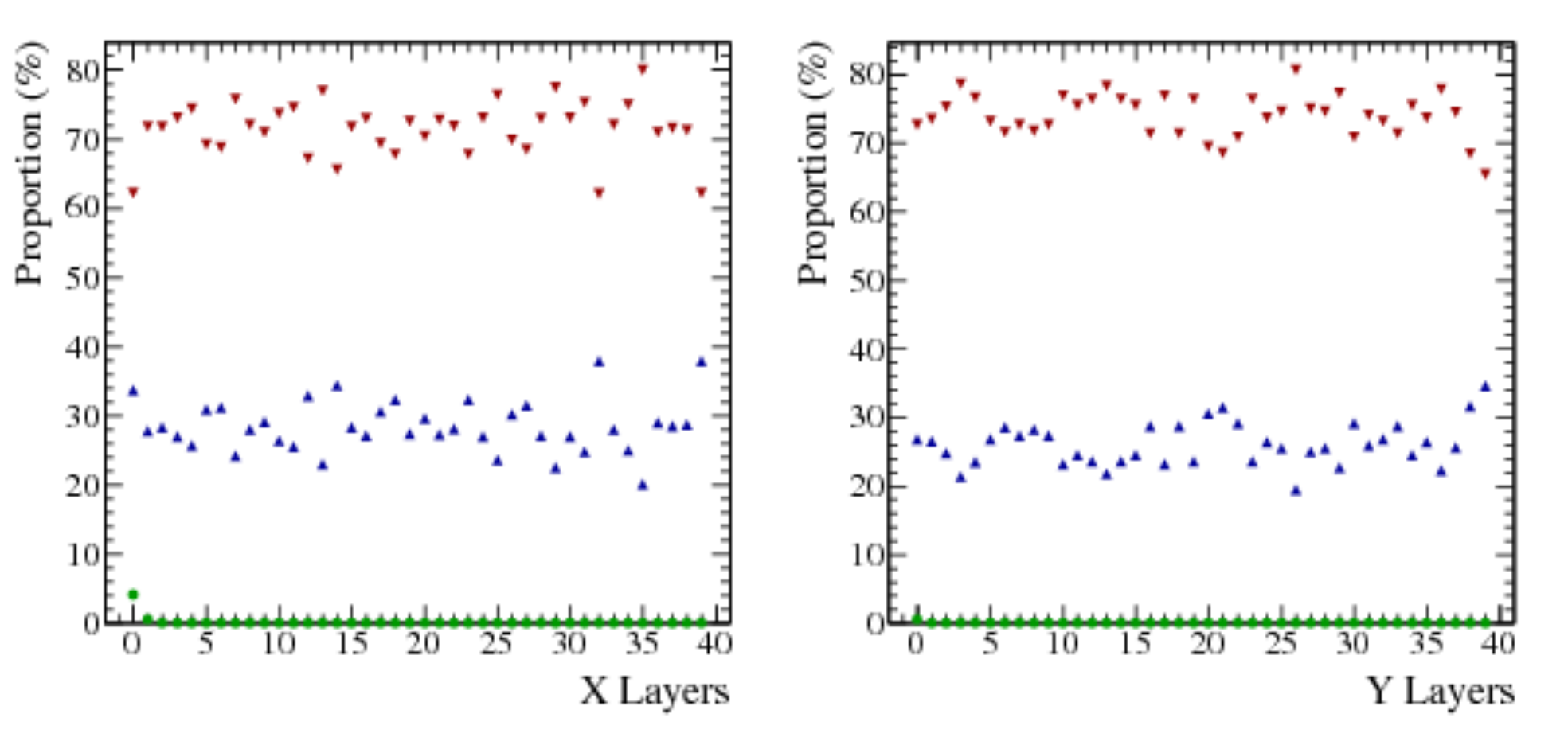}
\caption{The \pod{} layer detection  efficiency.  The plots show the
  proportion of  tracks with 0 (green circles),  1 (blue triangles) and 2  or more (red inverted triangles)
  hits in a  layer, for both $x$ and  $y$.  The small excess of  0 hits in
  the upstream $x$ layer is due  to neutrino interactions in the first $y$
  layer passing the cuts.}
\label{fig:layer-efficiency}
\end{figure}

The  results, shown  in  Fig.~\ref{fig:layer-efficiency},  
indicate the probability of finding 0, 1 or 2 hits in each $x$ or $y$ plane.
The  tracking  efficiency  is  100\%  for  all  but  the  first  three
scintillator  planes, which  is  explained by  the selection  criteria
allowing a small number of  first layer neutrino interactions into the
sample.

\begin{figure}[htb]
\centering
\includegraphics[width=0.9\linewidth]{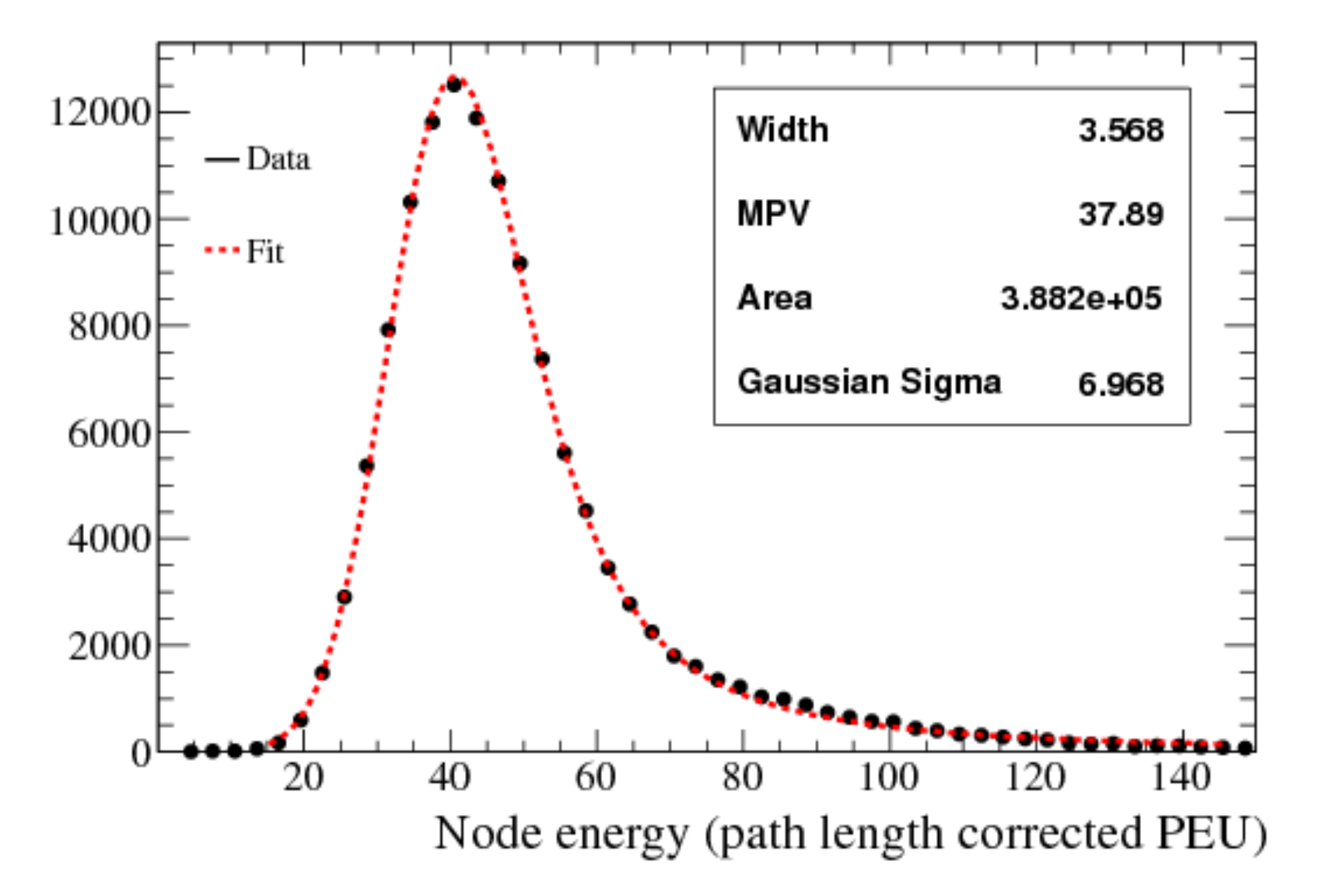}
\caption{The summed charge deposit for doublets from through-going muons originating from neutrino interactions outside the detector. The fit is a Gaussian-Landau distribution.}
\label{fig:mip-doubletlightyield}
\end{figure}

Figure~\ref{fig:mip-doubletlightyield} shows the summed charge deposit
for the  two hit sample,  after calibration and path  correction using the reconstructed track angle. The
plot  has been fit  with a Gaussian-Landau  distribution, and  returns a
most probable value of 37.9  p.e./mip/cm.  This value provides a known
point, which each channel of the \pod{} can be calibrated to, ensuring a
constant response for the detector.

\subsection{LIS Operation and Performance}

\begin{figure}[h]
\centering
\includegraphics[width=0.5\textwidth]{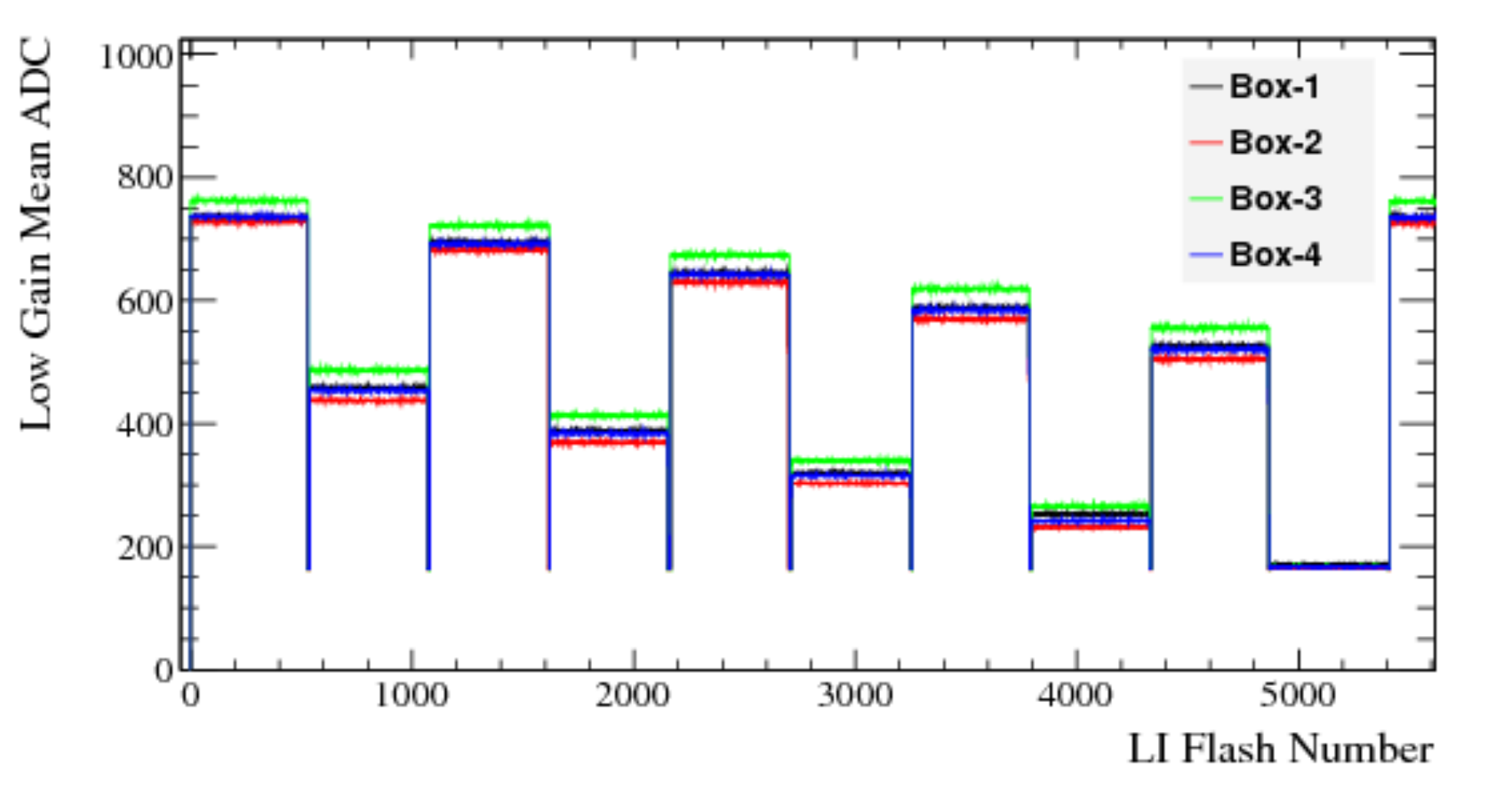}
\caption{Average signal produced by each pulser for a full cycle of 10 amplitudes.}
\label{figcycle}
\end{figure}

The LIS system simultaneously illuminates the entire \pod{} and is read out at in bursts of 20 Hz interspersed with other trigger types. 
The current settings give the LIS system an effective rate of 1.5Hz. The LIS system cycles through a set of ten amplitudes, each with 500 flashes, taking about one hour for a complete cycle. Figure~\ref{figcycle} shows the average ADC signal produced by each of the four pulser boxes during a typical run. 
Each plateau corresponds to a single amplitude. The sequence of amplitude was purposefully chosen to produce a clear step structure in the response to enable easy visual separation of the groups from each other. 

\begin{figure}[h]
\centering
\includegraphics[width=0.5\textwidth]{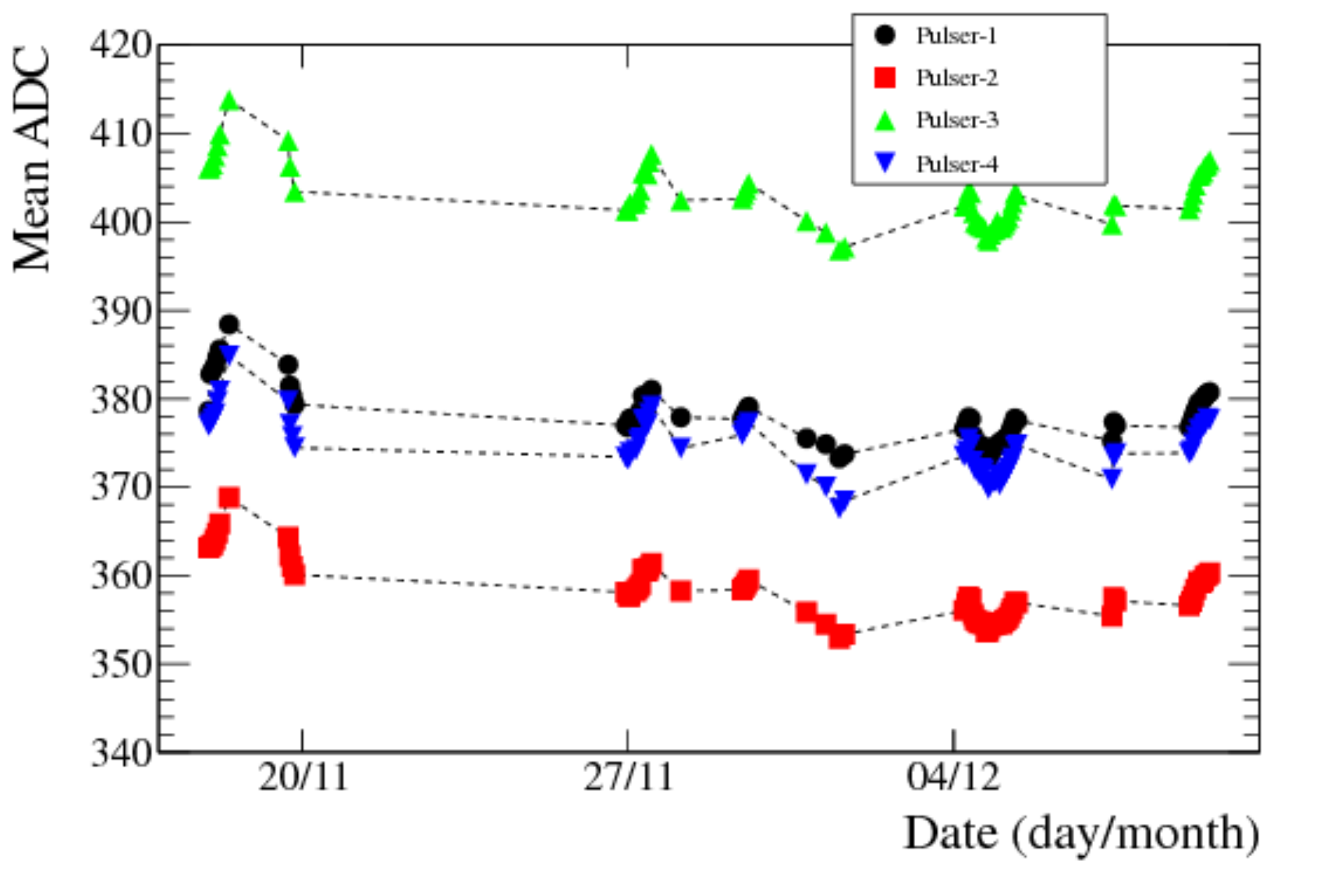}
\caption{\pod{} stability for a portion of a physics run showing the average signal in all photosensors illuminated by the LIS system.  Note the suppressed zero.}
\label{figstability}
\end{figure}

Besides providing the ability to quickly determine the correct functioning of all \pod{} photosensors, the LIS provides a tool to monitor the stability of the photosensor signal, shown for a portion of a physics run is shown in Fig.~\ref{figstability}.
The variation over short periods of time can be attributed to changes in the photosensor gain. Shifts that are different with respect to each 
pulser can be evidence for malfunctions in the \pod{} readout electronics. 

\subsection{Water Target Filling and Monitoring}

\begin{figure}[h]
\centering
\includegraphics[width=0.5\textwidth]{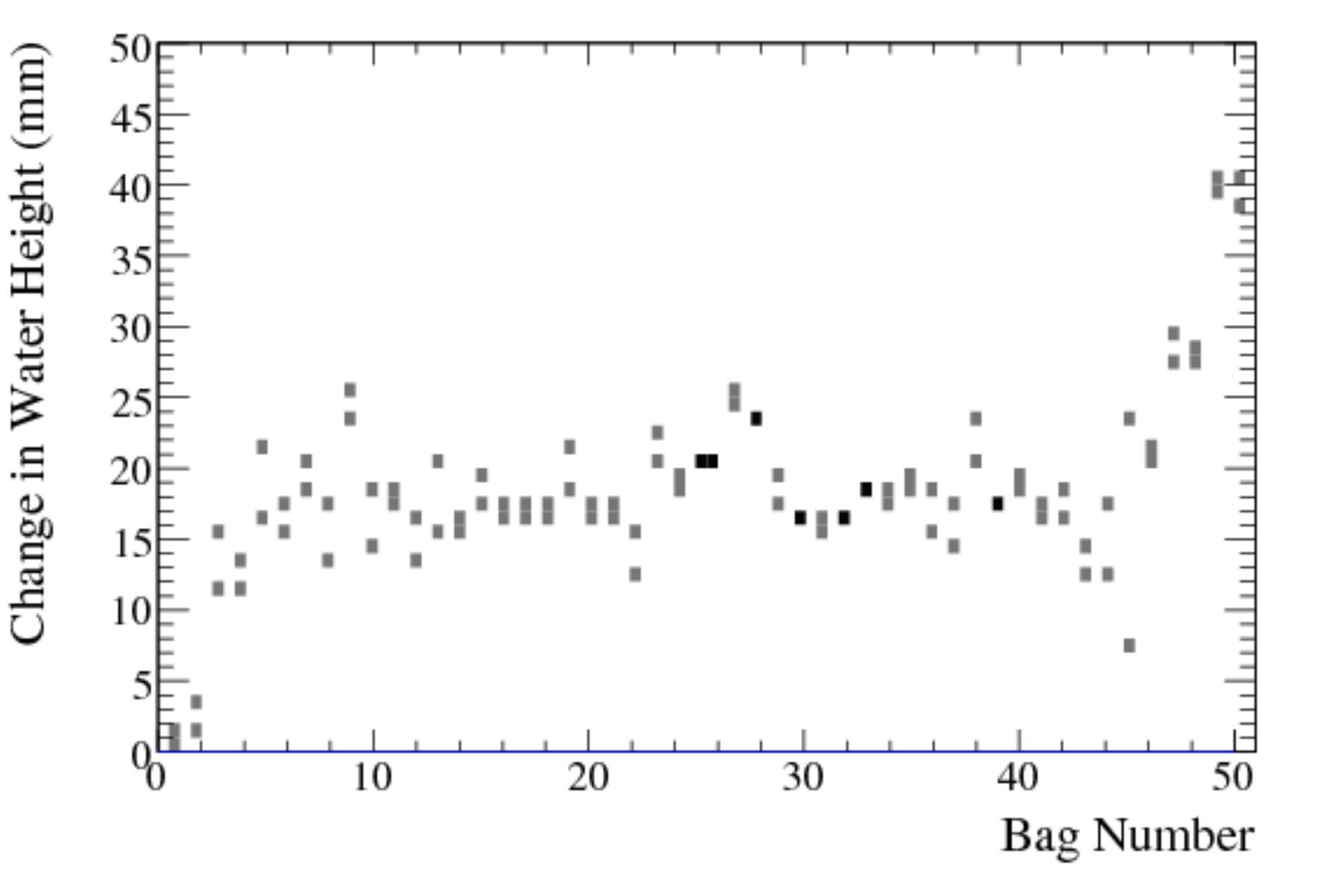}
\caption{Water level shift after 19 days (2 sensors per bag).} \label{waterperformance_figure1}
\end{figure}

The depth sensors were found to have fluctuations of \(\pm\)1 mm but had a \(\pm\)15 mm calibration offset before insertion into the \pod{}.  This offset was reduced by using the fixed binary wet-dry level sensors to provide calibration reference points \emph{in situ}. We expected the water level to drop in some layers due to deflection of the plastic scintillator.  As shown in Fig.~\ref{waterperformance_figure1}, the largest change in water level is closest to the downstream end of the \pod{} which is not directly supported by the basket.

Geometry and the measured dimensions of the \pod{} constrain the uncertainty on the total mass of water in the fiducial volume to approximately 3\%.  
The addition of measurements from the WL400 depth sensors and the external tank volume measurements reduce this uncertainty to less than 1\%.

\section{Conclusion}
\label{sec:conclusion}

The \pod{} subdetector in the T2K near detector will be used to meet the T2K physics goals by 
providing a determination of the neutral current $\pi^{0}$ rate for the T2K neutrino beam on a water target.  Analyses on this measurement and several others are underway.

The detector was installed in October 2009 and has been taking neutrino beam data since January 2010.  To date, there have been no major problems  and the \pod{} is performing as expected.

\section{Acknowledgements}
\label{sec:acknowledgements}

The \pod{} detector has been built and operated using funds provided by
the  U.S. Department of Energy.  In addition, the participation of individual researchers
and institutions in the construction of the \pod{} has been
further supported by funds from the U.S. Department of Energy Early Career Program and from the City University of New York PSC-CUNY Research Award Program.   The authors also wish to acknowledge the support provided by 
the collaborating institutions, particularly the State University of New York at Stony Brook, 
Office of the Vice President for Research and finally to thank our T2K colleagues for their invaluable help during the installation and commissioning of the detector.




\bibliographystyle{./elsarticle/model1a-num-names}
\bibliography{P0DNIM}







\end{document}